\documentclass[12pt,preprint]{aastex63}
\usepackage{graphicx}
\usepackage[]{xcolor}

\newcommand{\ts}[1]{\textcolor{blue}{#1}}

\received{...}
\revised{...}
\accepted{...}
\submitjournal{ApJ}

\begin{document}

\title{Observing the Time Evolution of the Multi-Component Nucleus of 3C\,84}

\correspondingauthor{Brian Punsly}
\email{brian.punsly@cox.net}

\author{Brian Punsly}
\affiliation{1415 Granvia Altamira, Palos Verdes Estates CA, USA 90274}
\affiliation{ICRANet, Piazza della Repubblica 10 Pescara 65100, Italy}
\affiliation{ICRA, Physics Department, University La Sapienza, Roma, Italy}

\author{Hiroshi Nagai}
\affiliation{National Astronomical Observatory of Japan, Osawa 2-21-1, Mitaka, Tokyo 181-8588, Japan}
\affiliation{The Graduate University for Advanced Studies, SOKENDAI, Osawa 2-21-1, Mitaka, Tokyo 181-8588, Japan}

\author{Tuomas Savolainen}
\affiliation{Aalto University Mets\"ahovi Radio Observatory, Mets\"ahovintie 114, 02540 Kylm\"al\"a, Finland}
\affiliation{Aalto University Department of Electronics and Nanoengineering, PL 15500, 00076 Aalto, Finland}
\affiliation{Max-Planck-Institut f\"ur Radioastronomie, Auf dem H\"ugel 69, 53121 Bonn, Germany}

\author{Monica Orienti}
\affiliation{INAF - Istituto di Radioastronomia, Via Gobetti 101, 40129 Bologna, Italy }

\begin{abstract}
The advent of global mm-band Very Long Baseline Interferometry (VLBI)
in recent years has finally revealed the morphology of the base of the
two most prominent nearby, bright, extragalactic radio jets in M\,87
and 3C\,84. The images are quite surprising considering the
predictions of jet theory and current numerical modeling. The jet
bases are extremely wide compared to expectations and the nucleus of
3C\,84 is very complicated. It appears as a double in 86\,GHz
observations with 50\,$\mu$as resolution and a triple nucleus with
30\,$\mu$as resolution with space-based VLBI by RadioAstron at
22\,GHz. What is even odder is that the double and triple are arranged
along an east-west line that is approximately orthogonal to the
north-south large scale jet on 150\,$\mu$as $-$ 4\,mas scales. We
explore the emergence of an (east-west) double nucleus in the lower
resolution 43\,GHz Very Long Baseline Array (VLBA) imaging from August
2018 to April 2020. The double is marginally resolved. We exploit the
east-west resolution associated with the longest baselines, $\sim
0.08$\,mas, to track a predominantly east-west separation speed of
$\approx 0.086\pm 0.008$\,c. We estimate that the observed mildly
relativistic speed persists over a de-projected distance of
$\sim 1900-9800$ times the central, supermassive black hole,
gravitational radius ($\sim 0.3-1.5$\,lt-yrs) from the point of
origin.
\end{abstract}


\section{Introduction}

The galaxy, NGC\,1275, harbors the radio source 3C\,84 with both a
parsec scale jet and a kpc scale jet as well as a large low frequency
radio halo \citep{ped90,wal00}. It has been the brightest
extragalactic radio source at high frequency with flux densities of
45\,Jy $-$ 65\,Jy in 1980 at 270\,GHz and $\sim 40-50$\,Jy at 90\,GHz
from 1965$-$1985 \citep{nes95}. The flare subsided in the 1990s, but a
new strong flare began in 2005 \citep{nag10,tri11}. Due to its
proximity ($z = 0.0176$), and its brightness, 3C\,84 along with M\,87
are the best candidates for resolving the jet near the launching
region with high frequency Very Long Baseline Interferometry
(VLBI). High resolution observations of M\,87 and 3C\,84 have
displayed extremely wide jet opening angles within 0.1\,mas of the
core, $\sim 127^{\circ}$ and $138^{\circ}$, respectively and both jets
are extremely edge brightened near the base
\citep{kim18,gio18,pun19}. The large degree of edge brightening and
the enormous (the maximum possible geometrically is $180^{\circ}$)
opening angles were not anticipated in seminal works on simple jet
theory \citep{bla79}. Numerical models of jet formation have been
designed to explain these extreme properties in M\,87. It was hoped
that choosing lines of sight (LOS) that are nearly aligned with the
jet axis and plasma emissivity/plasma enthalpy profiles (injected by
numerical researchers) that were specifically designed to produce the
high resolution radio images would resolve the conflict
\citep{mos16,cha19}. However, current numerical simulations still
produce synthetic jet images that are far too narrow and not nearly
edge brightened enough near their bases when compared directly to the
highest resolution interferometric images of the jet base
\citep{pun19}. This wide base seems to abruptly transition to a highly
collimated inner jet. The powerlaw fit to the jet width, $W(z)$, as a
function of axial displacement along the jet, $z$, is $W(z) \propto
z^{k}$, $k=0.230\pm 0.049$ for $0.06 \rm{mas} < z < 0.3 \rm{mas}$ in M87, even though the jet might be parabolic (k=0.5) at
larger z \citep{pun19}. A similar abrupt transition also seems to
occur in the jet of 3C 84 juxtaposed to its wide base, see
\citet{gio18} or Figures 9 and 10 of this paper. Furthermore, 3C\,84
was shown to have a triple nucleus in the highest resolution radio
observation, 30\,$\mu$as, performed with RadioAstron at
22\,GHz\footnote{We define the multiplicity of the nucleus,
    i.e. double or triple, as the apparent number of components that
    can be discerned from visual inspection of the image. This
    designation is a function of the resolution of the telescope. If
    the resolution were much higher then there could be many
    small components that could not be seen at lower resolution.}
\citep{gio18}. The triple is distributed primarily along the east-west
direction roughly orthogonal to the jet that is directed towards the
south and is defined by two bright ridges, beginning 150\,$\mu$as to
the south (see the schematic diagram in Figure~1). These circumstances
suggest that there is jet launching physics to be discovered with high
resolution of VLBI, and not just the verification of existing theory.

Thus motivated, this study explores the nature of the multi-component
nucleus that has been seen with the highest resolution VLBI, $\sim$
30\,$\mu$as in the east-west direction with RadioAstron on September
21-22, 2013 \citep{gio18}. Our method is to use the lower resolution
7\,mm Very Long Baseline Array (VLBA) data that is created for the
purpose of approximately monthly monitoring by a Boston University
based research effort, the VLBA-BU Blazar Monitoring
Program\footnote{http://www.bu.edu/blazars/VLBAproject.html}
\citep{jor17}. Thus, we can, in principle, detect time evolution of
the nuclear region. A partially resolved double nucleus has appeared
in the CLEAN images from August 26, 2018 to April 7, 2020. We rely on
the publicly available files from the BU website to extract the
highest resolution data, corresponding to the longest baselines
associated with the 43\,GHz VLBA.

\begin{figure}
\begin{center}
\includegraphics[width=120 mm, angle= 0]{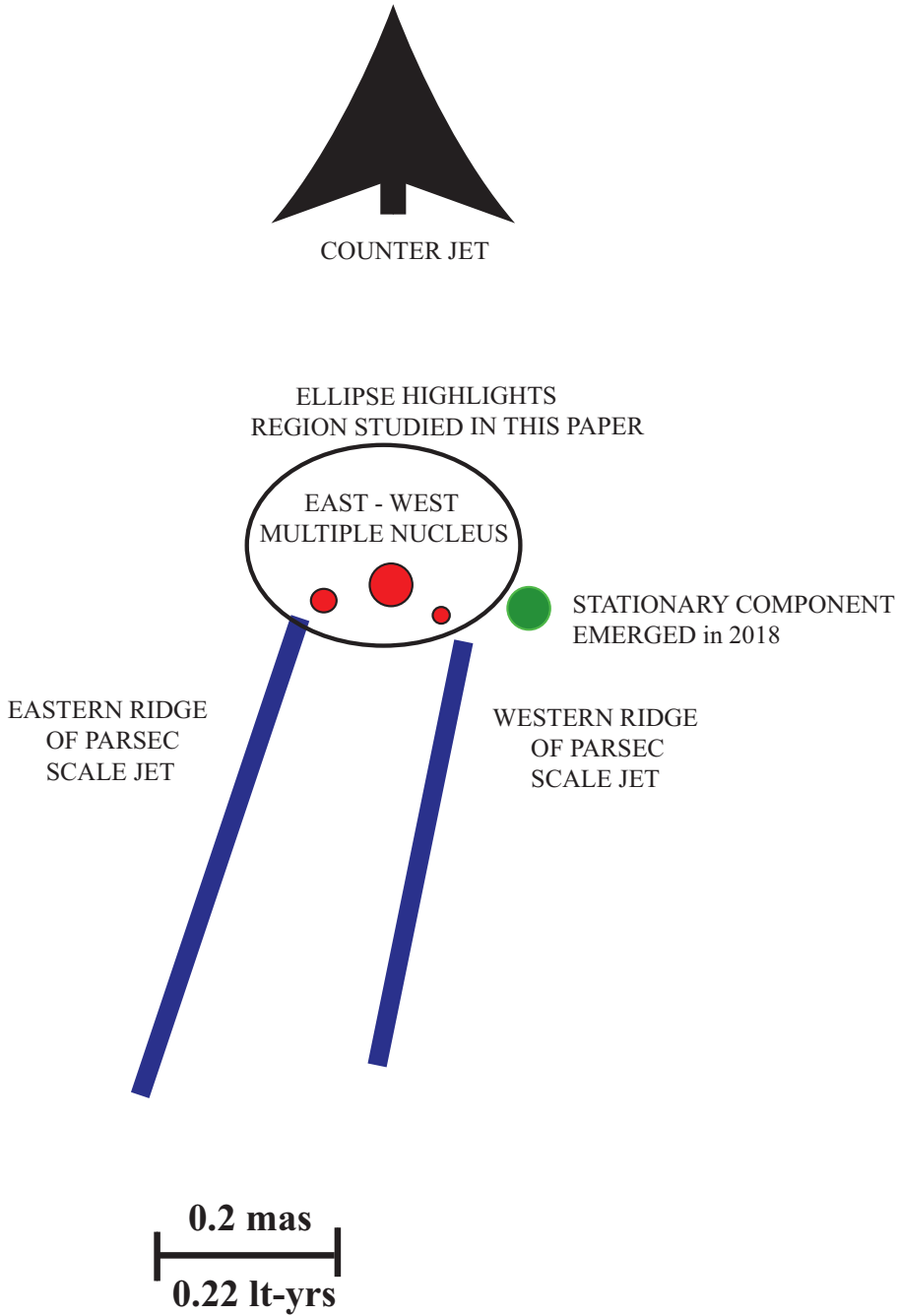}
\caption{The taxonomy of prominent features in the nuclear region of
  3C\,84. The position angle of the southerly directed jet swings back
  and forth from east to west by $\sim 35^{\circ}$ over periods of
  about 5-10 years.}
\end{center}
\end{figure}

We begin with a description of the very complex nuclear inner light
year of the jet that has been revealed in recent high resolution VLBI
campaigns. Figure~1 is a schematic diagram of the numerous nuclear
features that have been detected in this nearby very bright radio
source. A schematic is much clearer than overlaying numerous images
made at various frequencies. An approximate scale is indicated at the
bottom, both in angular size and physical dimension (projected on the
sky plane). The primary feature is the triple nucleus (enclosed by the
ellipse) from the 22\,GHz RadioAstron observations \citep{gio18}. This
region is the focus of this study. This nuclear feature is apparently
not time stationary. A more recent 86\,GHz global VLBI observation on
May 16, 2015 revealed this region to be a double nucleus along the
east-west direction \citep{kim19}. With 43\,GHz VLBA, a double nucleus
has never been identified previously. Based on the aforementioned
higher resolution VLBI, the east-west separation is $\sim
0.1-0.15$\,mas, similar to the nominal east-west restoring beam of
  the uniformly weighted 10 station VLBA of $0.15$\,mas. Thus, we
expect to see partially resolved structure in epochs of wide
separation.

Figure~1 shows other complicating features as well and helps to
illustrate the ultimate goal of finding a unifying picture that
incorporates all of these important elements. First, there is the
faint counter-jet that has been seen at numerous frequencies
\citep{wal00,lis01,fuj17}. Then there are the very prominent east and
west ridges that frame the jet headed to the south. They are of varied
prominence from epoch to epoch and observation to observation. In the
high resolution 22\,GHz RadioAstron observation they are significantly
brighter than the almost hollow interior of the jet. This ``double
rail" configuration has also been seen with 43\,GHz VLBA
\citep{nag14}. A similar double rail configuration has been detected
in high resolution VLBI images of M\,87 adjacent to the nucleus as
well \citep{had14,kim18}. The most curious aspect of these ridges in
the 43\,GHz BU VLBA monitoring data is that most of the time one ridge
is much brighter than the other for periods of time that last
years. The fundamental question that we wish to gain some insight into
is: what is the relationship between the multiple east-west nucleus
and these vacillating bright ridges of emission that emerge almost
orthogonal to the axis of nuclear emission, in the southern direction?

Some of these features are evident in the magnified views of the
nuclear region from the total intensity images taken from the BU
website in Figure~2. We note one other aspect of the nuclear region. A
modest feature has emerged to the west of the nucleus in 2018, about
0.25\,mas away. It appears to be real because it is persistent and is
seen a few contours above the noise level. It also seems to be
stationary. With all the components emerging in almost every
direction, the situation is not well explained by a simple pencil-beam
jet. We are actually getting close to the jet launching region, so we
are starting to uncover some of the messy details of the physical
mechanism. We do not attempt to resolve these complexities in the
present paper. We take a more modest approach and merely try to establish and
quantify the rate of nuclear expansion within the central ellipse of
Figure~1.

Section~2 describes our methods of extracting the nuclear separation
from the data. In Section~3, we describe how we use the CLEAN
component (CC) models from the BU program to extract the east-west
resolution associated with the longest baselines. Section~4 presents
our estimate of the apparent speed of separation of the double
nucleus. In Section~5, we explore the non-trivial methods of
  fitting the nuclear components in the $(u,v)$ plane using Gaussian
  models. In Section~6, we compare the nuclear configuration observed
by the 43\,GHz VLBA in 2018-2020 with the nuclear triple observed by
RadioAstron in 2013. Throughout this paper, we adopt the following
cosmological parameters: $H_{0}$=69.6 km s$^{-1}$ Mpc$^{-1}$,
$\Omega_{\Lambda}=0.714$ and $\Omega_{m}=0.286$ and use Ned Wright's
Javascript Cosmology Calculator website \citep{wri06}. In our adopted
cosmology we use a conversion of 0.360 pc to 1 mas.

\begin{figure*}
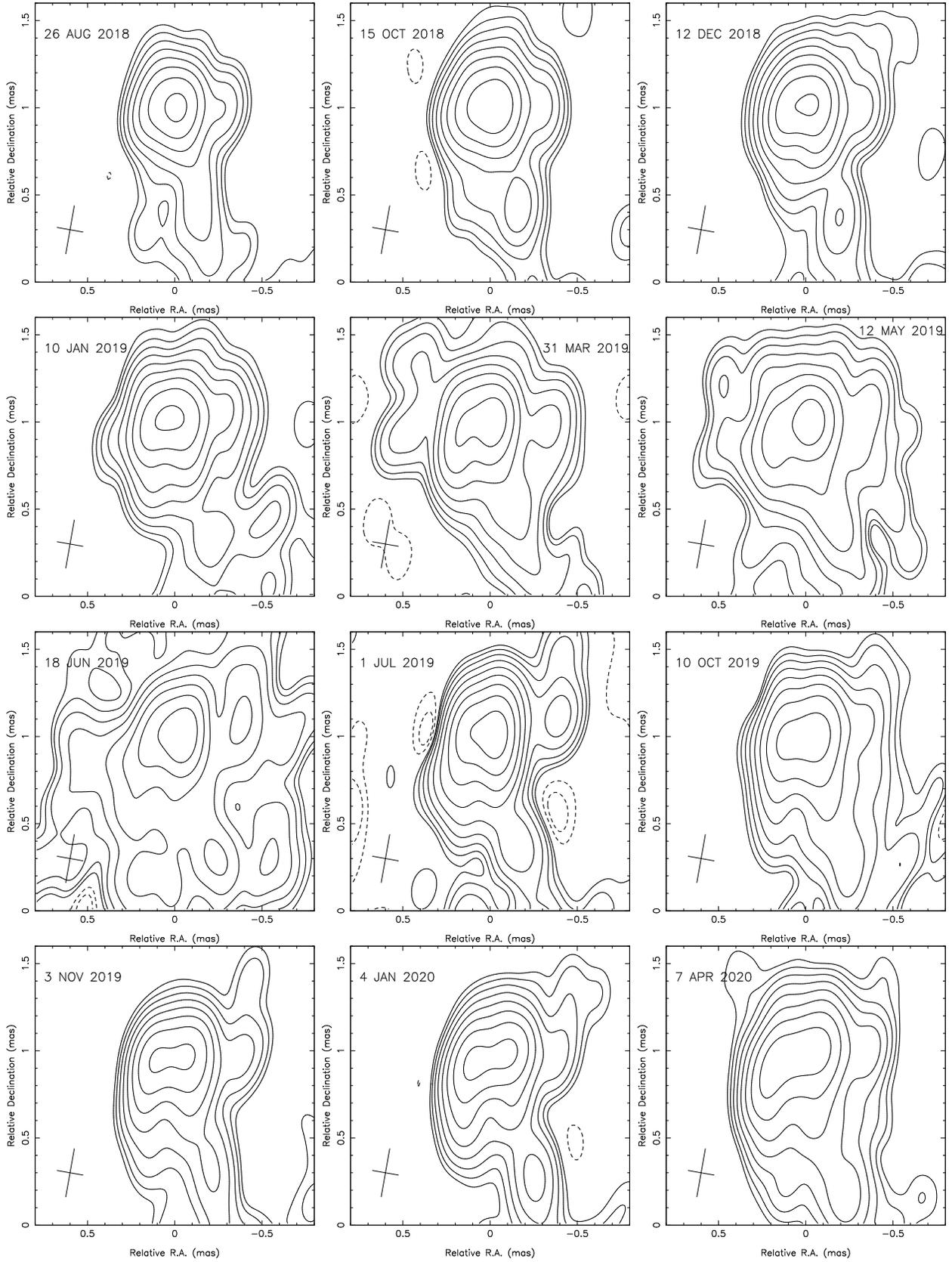

\begin{center}
\includegraphics[width=55 mm, angle=-90]{f2.eps}
\includegraphics[width=55 mm, angle= -90]{f3.eps}
\includegraphics[width=55 mm, angle= -90]{f4.eps}\\
\includegraphics[width=55 mm, angle= -90]{f5.eps}
\includegraphics[width=55 mm, angle= -90]{f6.eps}
\includegraphics[width=55 mm, angle= -90]{f7.eps}\\
\includegraphics[width=55 mm, angle= -90]{f8.eps}
\includegraphics[width=55 mm, angle= -90]{f9.eps}
\includegraphics[width=55 mm, angle= -90]{f10.eps}\\
\includegraphics[width=55 mm, angle= -90]{f11.eps}
\includegraphics[width=55 mm, angle= -90]{f12.eps}
\includegraphics[width=55 mm, angle= -90]{f13.eps}
\caption{\footnotesize{The nucleus in the 43\,GHz image starts
    elongating in the east-west direction in August 2018. The nucleus
    appears almost fully resolved by November 2019. The eastern
    component seems to veer towards the large scale southerly directed
    jet in the 2020 images. The details of the degree of degradation
    of many compromised observations in 2019 can be found in Table
    1. The June and July 2019 images reflect the significant
    degradation indicated in Table 1. The contours start from 10
    mJy/beam and increase in steps of 2x in all the plots. The
    restoring beam is indicated by the cross in the lower left.} }
\end{center}
\end{figure*}

\begin{figure*}
\begin{center}
\includegraphics[width=40 mm, angle= 0]{f14.eps}
\includegraphics[width=40 mm, angle= 0]{f15.eps}
\includegraphics[width=40 mm, angle= 0]{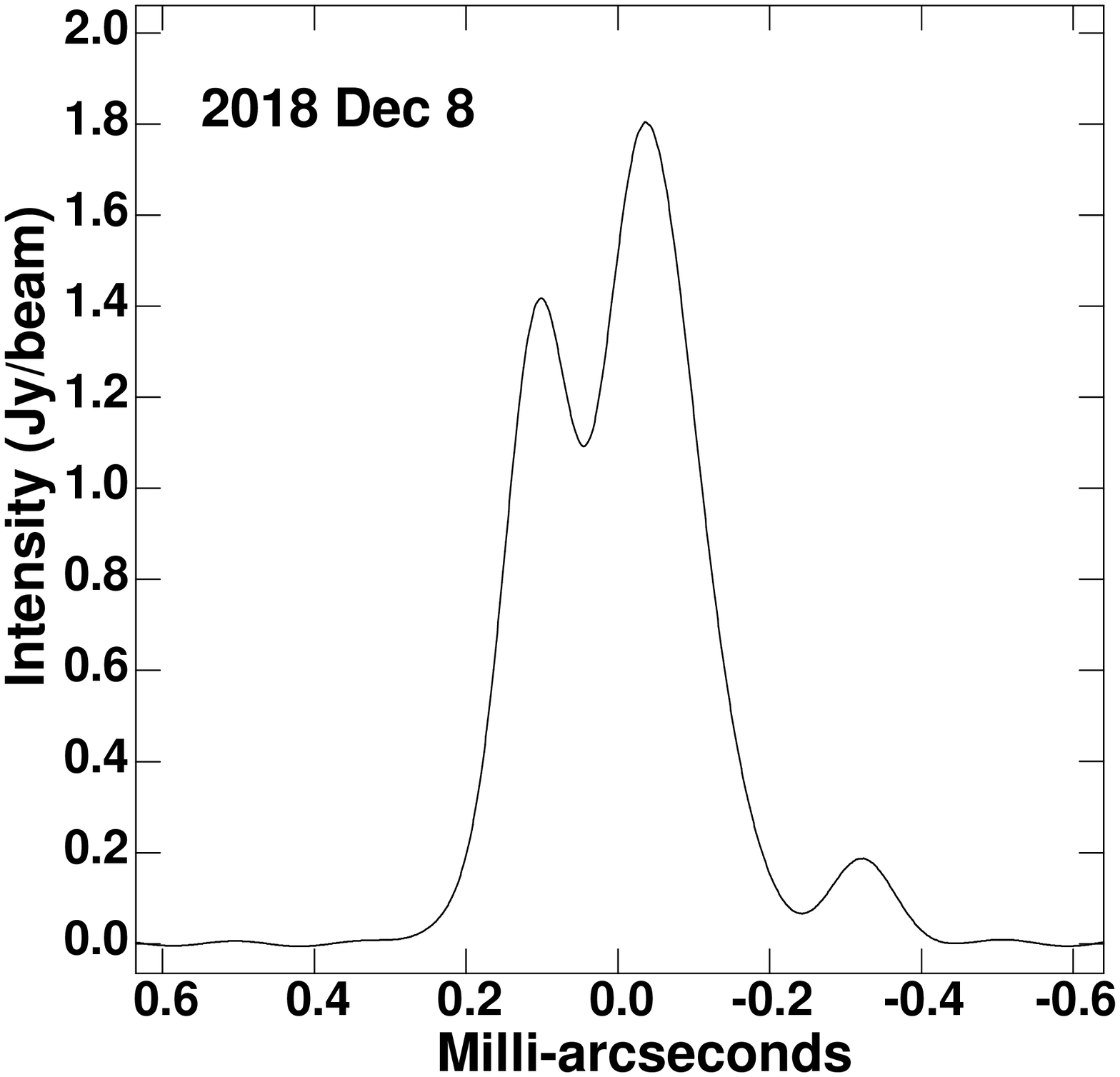}
\includegraphics[width=40 mm, angle= 0]{f17.eps}
\includegraphics[width=40 mm, angle= 0]{f18.eps}
\includegraphics[width=40 mm, angle= 0]{f19.eps}
\includegraphics[width=40 mm, angle= 0]{f20.eps}
\includegraphics[width=40 mm, angle= 0]{f21.eps}
\includegraphics[width=40 mm, angle= 0]{f22.eps}
\includegraphics[width=40 mm, angle= 0]{f23.eps}
\includegraphics[width=40 mm, angle= 0]{f24.eps}
\includegraphics[width=40 mm, angle= 0]{f25.eps}
\caption{\footnotesize{East-west intensity cross sections created from
    the CLEAN images. They are fixed in the north-south direction by
    locating them to pass through the peak intensity. Notice the
    emergence of a local maximum to the east of the peak on August 26,
    2018. A double peak is clearly seen on October 15, 2018. A third
    local maximum emerges in October 2019. Notice that January 10,
    2019 has a different cross-section with just a single peak. This
    epoch requires special attention (see Appendix B).}}
\end{center}
\end{figure*}

\begin{figure}
\begin{center}
\includegraphics[width=85 mm, angle= 0]{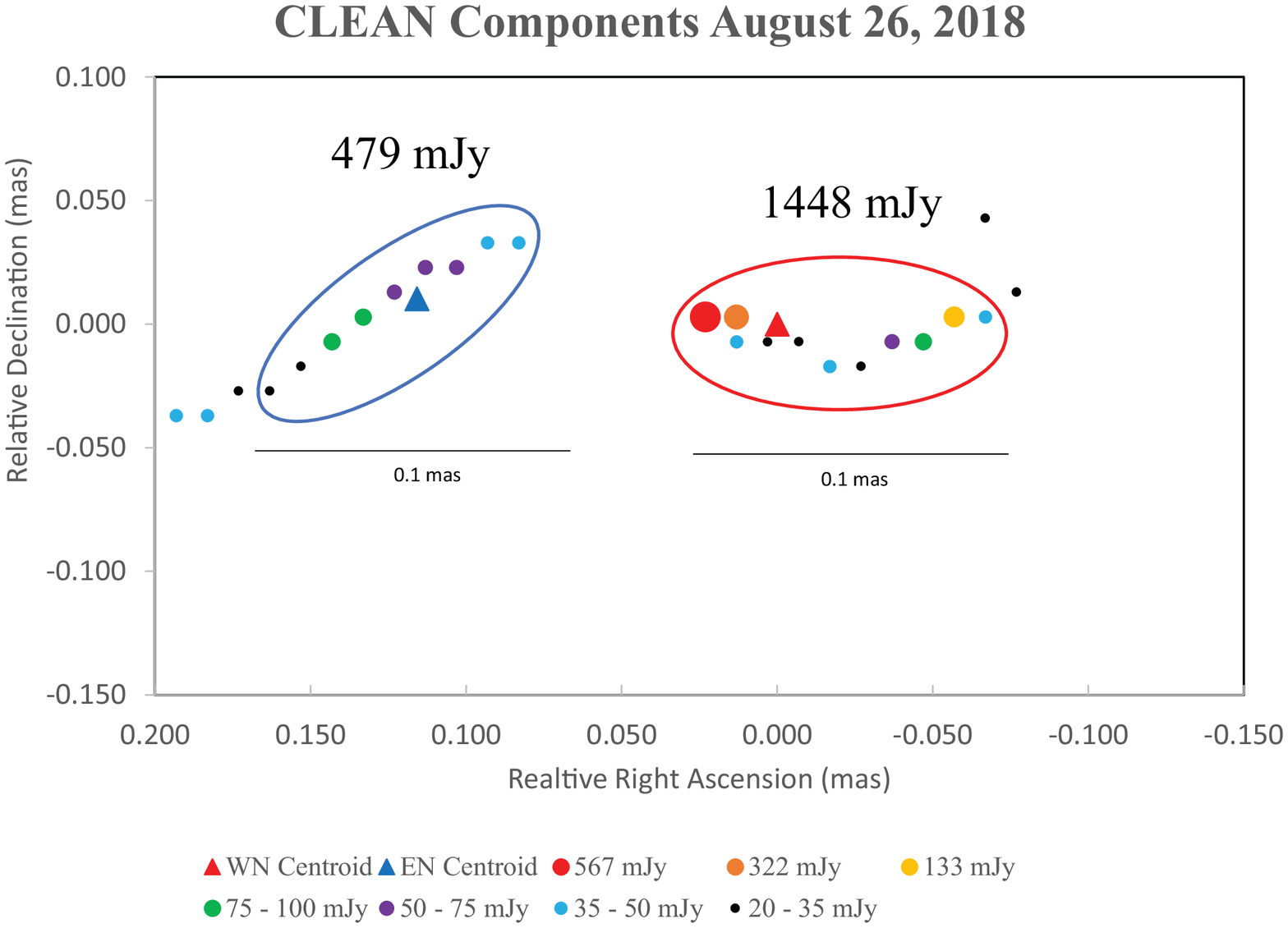}
\includegraphics[width=85 mm, angle= 0]{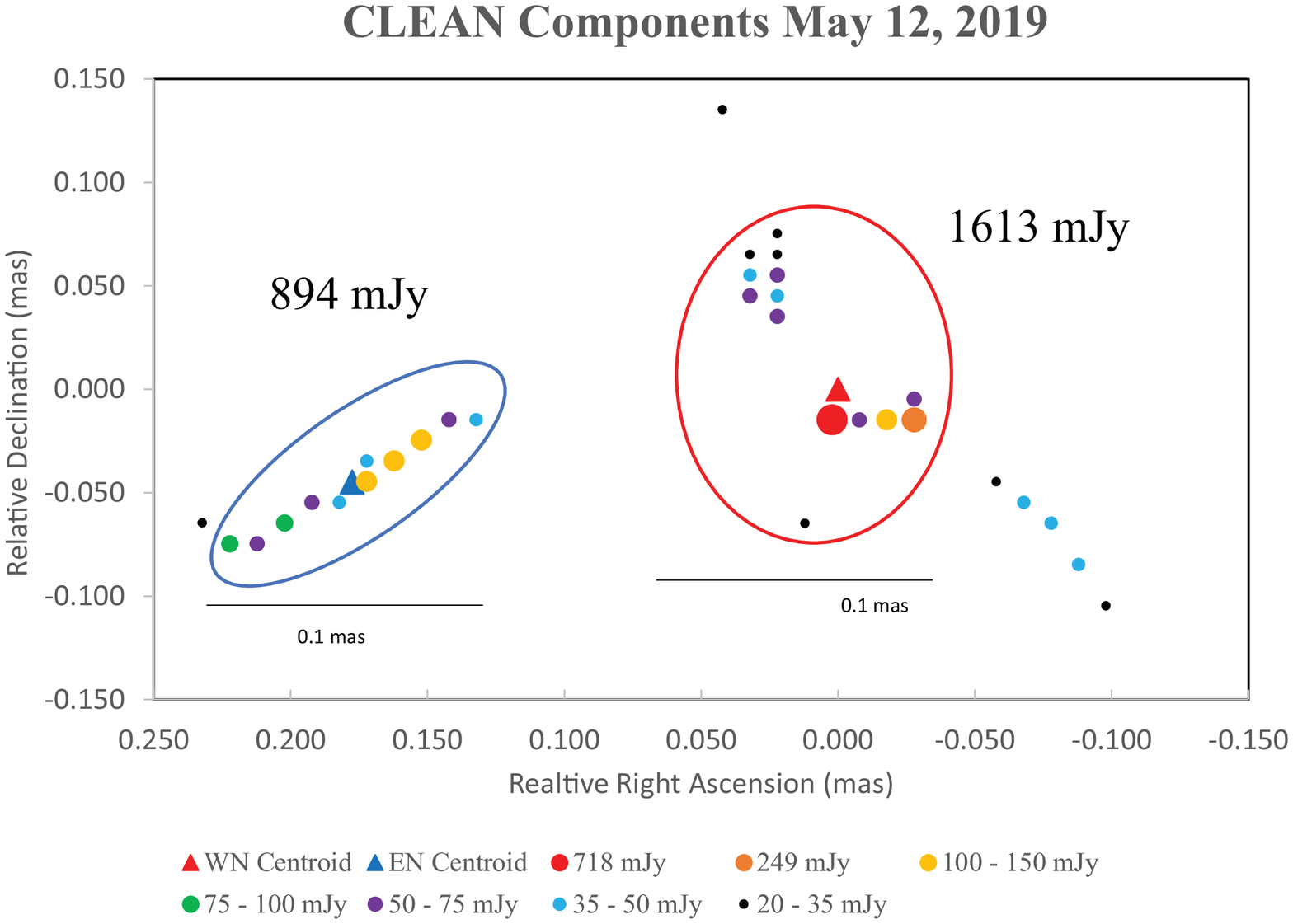}
\includegraphics[width=85 mm, angle= 0]{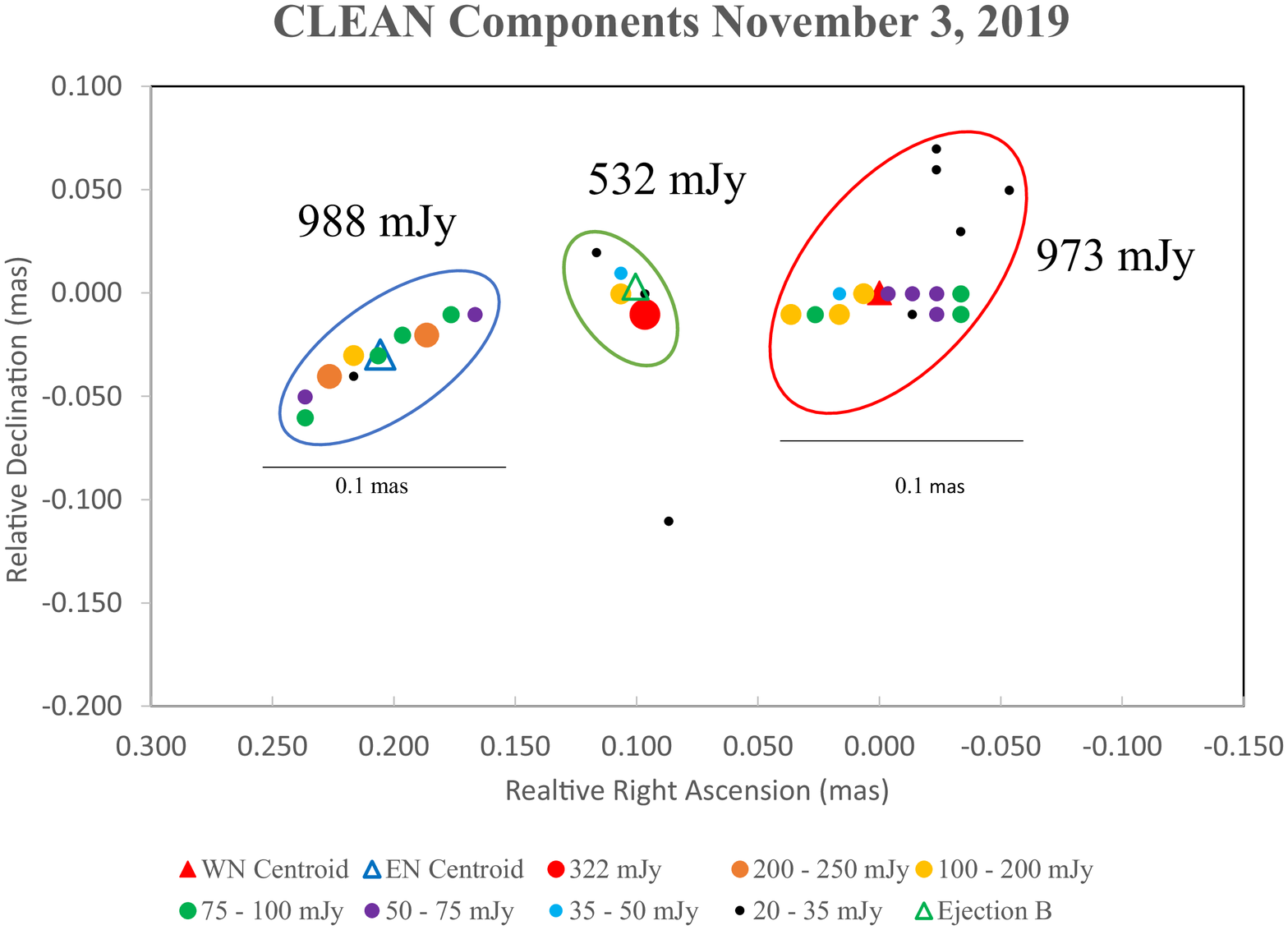}
\includegraphics[width=85 mm, angle= 0]{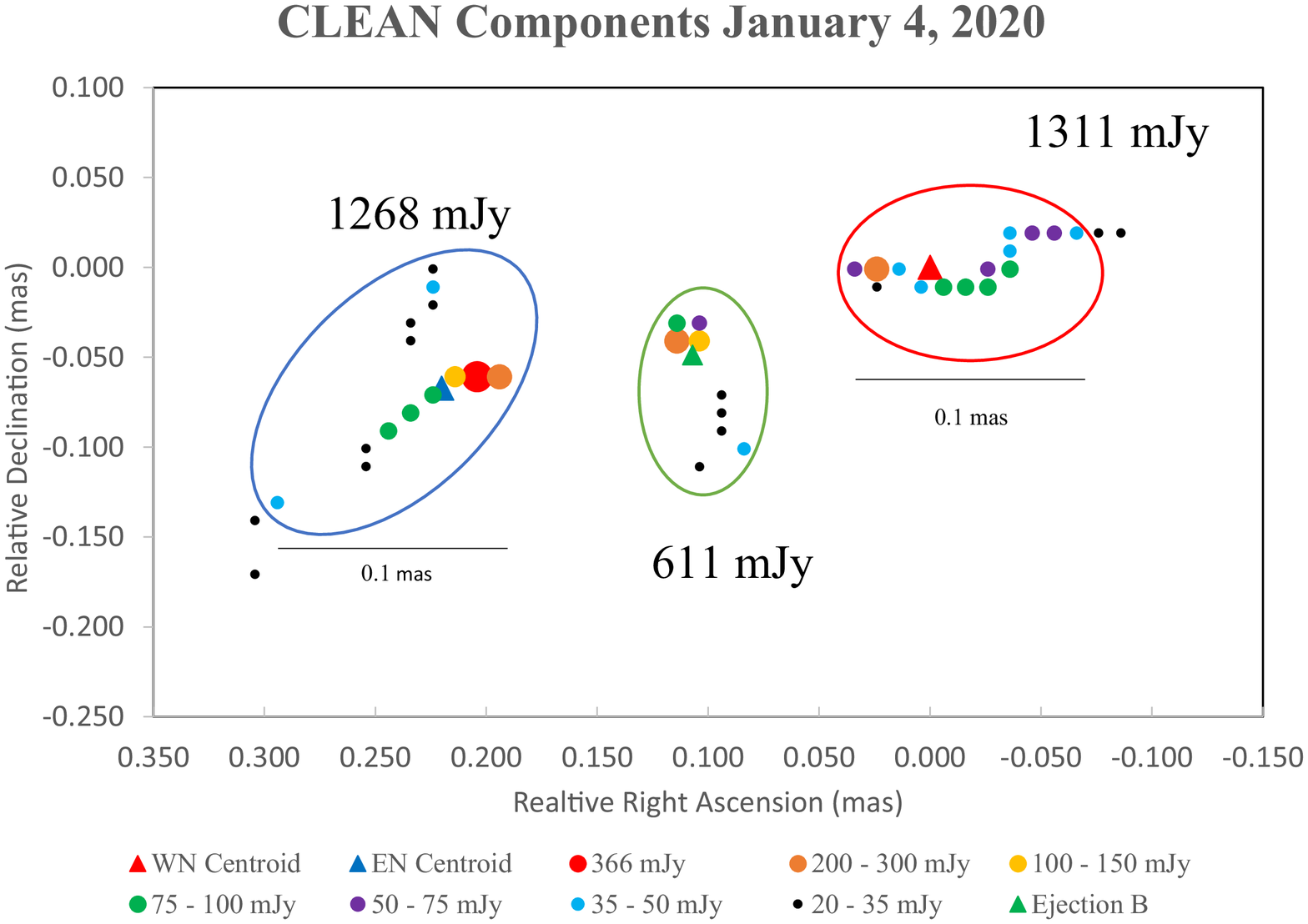}
\caption{The location of the nuclear CCs in four epochs. The two
  examples in the top row are used to illustrate our methods of
  defining which CCs determine the WN and EN. The WN and EN are
  defined by the CCs enclosed within red and blue ellipses,
  respectively. The red and blue triangles are the
  corresponding centroids of the flux density for each physical
  component. The details of these constructions are provided in the
  text. Note that the CCs outside the solid ellipses are weak
  (20$-$35\,mJy) and do not affect the centroid estimates, but we cut
  the CC distribution off once it approaches 0.1\,mas in east-west
  width as motivated in the text. The two examples in the bottom row
  show the emergence of a weaker (green) component in the fall of
  2019. We consider EN as the first ejection, ejection A, and the
  green cluster of CCs as a second ejection, ejection B.}
\end{center}
\end{figure}

\section{Method of Estimating Nuclear Expansion}

It is fortuitous that the nucleus expands primarily in the east-west
direction, this is also the direction of maximum resolution of the
VLBA at 43\,GHz (the major axis of the restoring beam has a position
angle of $-10^{\circ}$). The full width at half maximum (FWHM) of the
beam is approximately 0.28\,mas\,$\times$\,0.15\,mas for uniformly
weighted $(u,v)$ data.  However, when the data have sufficiently high
signal-to-noise ratio, the interferometer can resolve structures
significantly smaller than the nominal beam size. In our case the
longest baselines provide a typical resolution of approximately
0.08\,mas in the east-west direction. The resolution of the
interferometer is defined here as the FWHM of a circular Gaussian
brightness distribution that gives a visibility amplitude of 50\%
times the zero-baseline value. Mathematically, this is expressed as
$\rm{resolution} = (91 M\lambda/$baseline-length)\,mas
\citep{mar85}. It is known that higher resolution can be
  achieved by analyzing the visibility domain and these resolution
  limits depend on the noise in the $(u,v)$ plane \citep{mar12}. However,
  working in the image plane has the advantage of less ambiguity as to
  whether the model chosen for the source is appropriate
  \citep{fom99}. Since we are measuring small changes to a small
  separation on a complex background, the ambiguities in the models
  can cause a significant degradation of accuracy. In Appendix A, we provide
  simulations in the visibility domain to motivate the \citet{mar85}
  resolution limit as appropriate to this study. These simulations incorporate all the details of the observations that degrade the very high resolution limits achievable with model-fitting the visibility data with near perfect circumstances. We incorporate the $(u,v)$ coverage, SNR, and the complexity of the source structure.
  Even though this analysis is in the $(u,v)$ plane,``analysis of a properly made CLEAN image via good analysis techniques should produce values and errors which are equal to those of model fitting visibility data"\citep{fom99}. The CLEAN component based method utilized in this paper will be shown to be an example of a ``good analysis technique" in the image plane. We are able to find Gaussian fitting methods in Section 5 that yield similar resolution limits to our CLEAN component based methods. This forms a bridge between the simulations in Appendix A and the CLEAN component based method presented here.

So, in theory, there is information within the observations that can
resolve east-west features larger than 0.1\,mas, provided their
signal-to-noise ratio (SNR) is high. The 0.28\,mas\,$\times$\,0.15\,mas uniformly weighted restoring beam is
  too large to do this cleanly, the resultant blurring prevents a
clear east-west determination. The images in Figure~2 show this to be
the case as there appears to be two heavily blurred components that
are separating over time. We need to extract these partially resolved
structures from the observations in order to quantify the east-west
separation over time. There are three possible methods, super-resolved
images based on a restoring beam smaller than the central peak of the conventional interferometer beam, direct analysis of the
clean component (CC) model, and fitting the visibilities with simple
models of the core structure. The super-resolved images highlight the
multiple nucleus, but create large artifacts at angles more than
$45^{\circ}$ from the east-west direction. The net result is a
suggestive image, but Gaussian fits to the multiple components fail to
converge to a unique decomposition. The same circumstance exists with
efforts to fit the visibilities with simple Gaussian models (see
Section 5). The other alternative is to use the CC models. Even though
the models cannot be considered robust in a single epoch, they should
be suitable for identifying a trend that persists over time,
especially with regards to very bright features as is the case
here. This is the primary guiding principle of the following analysis.

\begin{deluxetable*}{cccccc}
\tablecaption{Compromised Observations in 2019}
\tabletypesize{\small}
\tablehead{\colhead{Date} &  \colhead{Missing Stations} & \colhead{Severely Compromised Stations} & \colhead{Comments} & \colhead{Quality}}
\startdata
02/03/2019 & MK and BR & ... & east-west resolution severely degraded & un-useable \\
02/08/2019 & NL & BR and HN much data deleted  & only $\sim 7$ stations disperses the  & un-useable \\
 &  & due to bad weather & distribution of CCs in image plane &  \\
03/31/2019 & HN & LA missing half of the data & good east-west resolution & useable  \\
 &  &  & from SC and MK, low SNR &   \\
05/12/2019 & SC & ... & MK and HN still provided & useable \\
 &  & ... & good east-west resolution &  \\
06/18/2019 & SC & ... & inaccurate pointing of & useable \\
 &  & ... &  the antennae, low SNR &  \\
07/01/2019 & SC & ... & inaccurate pointing of  & useable \\
 &  & ... &  the antennae, low SNR &  \\
10/19/2019 & BR & ... & low SNR & useable \\
\enddata
\tablecomments{Details of the observations generously provide by
  A. Marscher, 2019. Station acronyms: MK (Mauna Kea), BR (Brewster,
  Washington), NL (North Liberty, Iowa), HN (Hancock, New Hampshire),
  LA (Los Alamos, New Mexico), SC (St. Croix)}
\end{deluxetable*}

Before describing our models we first note that most observations in
2019 are degraded to varying degrees by bad weather, down observing
stations and inaccurate antennae pointing. The issues are summarized
in Table 1. Modest degradation of observations in 2019 adversely
affect the CC models and one of the main tasks of this analysis is to
quantify this degradation in terms of an uncertainty in our results.

In order to transition from the images in Figure~2 to the CC models,
we investigate intensity profiles of the CLEAN image along a
particular direction (east-west). We define a pixel size as
0.005\,mas\,$\times$\,0.005\,mas. We center an array of pixels on the
peak of the intensity. We place a circular Gaussian beam with a
0.1\,mas FWHM at the location of peak intensity and measure the
intensity. This is about 25\% larger angular resolution than the
angular resolution of the longest baselines of the array. We then
slide the circular Gaussian beam to the east and the west in discrete
0.005\,mas increments in order to generate the intensity
cross-sections in Figure~3. The intensity profiles were generously
provided by Alan Marscher. There is a clear local maximum to the east
of the peak on August 26, 2018. This corresponds to the slight
elongation towards the east in the first panel of Figure~2. A double
peak is clearly seen on October 15, 2018 and December 9, 2018 which
appears as increasing elongation of the nucleus towards the east in
Figure~2. The spacing between the peaks keeps increasing and on March
31, 2019 and May 12, 2019 there is clearly two distinct components in
the image plane. Notice that January 10, 2019 has a different
cross-section with just a single peak. This is out of family with the
other epochs, including the epoch just before and the epoch just after
in December and in March, respectively. Consequently, we analyze this
observation separately in Appendix B in order to assess the reason
that the uniform cross-sectional procedure failed to pick up the
second peak and nature of the root cause.

\section{The CLEAN Component Models}

CC models are provided for each observation epoch on the BU
website. During the first year of monitoring in Figure 2, the western
component of the double nucleus (WN, hereafter) is brighter and
fortunately is predominantly a tight east-west distribution of the
bright CCs that always includes the brightest CC. We face two primary
technical issues that we illustrate in Figure~4 with four
examples. The issues are:
\begin{enumerate}
\item How should the cluster of components be defined for both the
  eastern component of the nucleus (EN, hereafter) and the WN?
\item How do we define the coordinates of the cluster of components
  and the uncertainty in this location?
\end{enumerate}
We explore these issue below.

\begin{deluxetable*}{cccccccccccc}
\tablecaption{Details of the CLEAN Component Models of the Double Nucleus}
\tabletypesize{\tiny}
\tablehead{
  \colhead{(1)} &  \colhead{(2)} & \colhead{(3)}  & \colhead{(4)} & \colhead{(5)} & \colhead{(6)}
  & \colhead{(7)} & \colhead{(8)} & \colhead{(9)} & \colhead{(10)}  & \colhead{(11)}  \\
  \colhead{Date} &  \colhead{Day} & \colhead{$F_{\nu, \mathrm{WN}}$} & \colhead{$F_{\nu,\mathrm{EN}}$} &
  \colhead{Uniformly Weighted Beam} & \colhead{$I_\mathrm{peak}$} & \colhead{Min.\,Neg.}
  & \colhead{Smallest} & \colhead{$\Delta \rm{RA}$} & \colhead{$\Delta \rm{dec}$}
  & \colhead{Separation} \\
   &   &   &  & \colhead{(Naturally Weighted Beam)} & & &  &  &   &   \\
  \colhead{} &  \colhead{} & \colhead{} & \colhead{} &
  \colhead{Size/PA} & \colhead{} & \colhead{Cont.}
  & \colhead{CC} & \colhead{} & \colhead{} & \colhead{} \\
  \colhead{} & \colhead{} & \colhead{(Jy)} & \colhead{(Jy)} & \colhead{(mas/degrees)} & \colhead{(Jy/beam)} &
  \colhead{(mJy/beam)} & \colhead{(mJy)} & \colhead{(mas)} & \colhead{(mas)} & \colhead{(mas)}
}
\startdata
08/26/2018 & 0 & 1.448 & 0.479 &  0.303 x 0.179/+11.0 & 1.572 &-15  &20 & $0.116 \pm 0.013$& $0.010\pm 0.021$ & $0.116 \pm 0.013$ \\
  &   &   &  & (0.370 x 0.231/+13.3) & & &  &  &   &   \\
10/15/2018 & 50 & 1.856 & 0.838 &  0.314 x 0.154/+3.6 & 2.462  & -14 &25 & $0.133 \pm 0.011$ & $0.001\pm 0.022$ & $0.133 \pm 0.011$ \\
  &   &   &  & (0.352 x 0.175/+1.1) & & &  &  &   &   \\
12/09/2018 & 105 & 2.707 & 1.441 & 0.345 x 0.166/+14.9 & 2.875 & -10  & 20 & $0.144 \pm 0.013$ & $-0.030\pm 0.024$ & $0.147 \pm 0.014$ \\
  &   &   &  & (0.381 x 0.190/+10.3) & & &  &  &   &   \\
01/10/2019 & 137 & 3.349 & 1.079 & 0.361 x 0.160/+21.4 & 2.974 & -15  & 20 & $0.147 \pm 0.014$ & $-0.075\pm 0.024$ & $0.165 \pm 0.017$ \\
  &   &   &  & (0.405 x 0.187/+21.0) & & &  &  &   &   \\
03/31/2019 & 217 & 1.325 & 0.901 &  0.306 x 0.153/+8.9  & 2.204  & -24 & 25 & $0.175 \pm 0.011$ & $-0.031\pm 0.021$ & $0.178 \pm 0.012$ \\
  &   &   &  & (0.344 x 0.185/+7.4) & & &  &  &   &   \\
05/12/2019 & 259 & 1.356 & 0.894 &  0.407 x 0.176/-13.2 & 2.064 & -17 & 20  & $0.178 \pm 0.014$ & $-0.045\pm 0.028$ & $0.183 \pm 0.015$\\
  &   &   &  & (0.435 x 0.197/-10.8) & & &  &  &   &   \\
06/18/2019 & 296 & 2.337 & 0.799 &  0.397 x 0.183/-16.4 & 2.462 & -29 & 35  & $0.173 \pm 0.015$ & $0.003 \pm 0.027$ & $0.173 \pm 0.015$\\
  &   &   &  & (0.432 x 0.211/-14.5) & & &  &  &   &   \\
07/01/2019 & 309 & 3.448 & 2.116 &  0.425 x 0.173/-14.3 & 4.030 & -36 & 35  & $0.149 \pm 0.014$ & $0.019 \pm 0.029$ & $0.150 \pm 0.014$\\
  &   &   &  & (0.434 x 0.210/-12.9) & & &  &  &   &   \\
10/19/2019 & 419 & 0.793 & 0.823 &  0.329 x 0.142/-5.2 & 1.840 & -29 & 35  & $0.193 \pm 0.010$ & $-0.043\pm 0.023$& $0.198 \pm 0.011$\\
  &   &   &  & (0.386 x 0.159/-8.3) & & &  &  &   &   \\
11/03/2019 & 434 & 0.973 & 0.988 &  0.318 x 0.153/+3.6 & 1.490 & -13 & 20  & $0.206 \pm 0.011$ & $-0.030\pm 0.022$ & $0.208 \pm 0.011$\\
  &   &   &  & (0.387 x 0.183/+0.6) & & &  &  &   &   \\
01/04/2020 & 496 & 1.311 & 1.268 & 0.330 x 0.165/+2.1 & 1.633 & -13 & 20  & $0.220 \pm 0.012$ &$-0.067\pm 0.023$ & $0.230 \pm 0.013$\\
  &   &   &  & (0.399 x 0.205/+0.1) & & &  &  &   &   \\
04/07/2020 & 590 & 1.634 & 1.613 & 0.316 x 0.144/+4.5 & 2.127 & -10 & 20  & $0.238 \pm 0.010$ & $-0.068\pm 0.022$ & $0.248 \pm 0.012$\\
  &   &   &  & (0.342 x 0.160/+6.4) & & &  &  &   &
\enddata
\end{deluxetable*}

\subsection{Defining the Clustering of the CLEAN Components}

This is a slightly subjective exercise at times in which we use the
visual evidence from Figures~2 and 3 and the structure of the
components in the previous epoch and the subsequent epoch. The peak
surface brightness is about $1.5 - 3.0$\,Jy/beam in each image. We
consider any CC with a flux density larger than $\sim 20-35$\,mJy as
significant. This varies depending on the quality of the
observation. This choice typically scales with the magnitude of the
strongest negative contour in the FITS image file (see Table~2). In
practice, components less than 40\,mJy never contribute significantly
to our flux density related results (component centroids and
uncertainties). Figure~4 plots all the components larger than 20\,mJy
in the vicinity of the nucleus in four epochs. The red ellipse defines
all the components that we associate with the WN and the blue ellipse
encapsulates all the components associated with the EN. We
  crudely estimate that the east-west spread of the CCs in the WN and
  EN will be distributed within a length $< 0.1$\,mas. This follows
  from de-convolving the 0.1\,mas FWHM restoring beam from the FWHM of
  the intensity cross-sections peaks in Figure~3. We get 20 estimates
  of the FWHM of peaks that range from 0.023\,mas to 0.112\,mas with a
  mean of 0.076\,mas. Truncating the east-west width of the CC
  distribution $< $0.1\,mas is the fundamental constraint for
  estimating which components should be included in the WN and
  EN. This approximate limit guarantees that the size of the
  distributions of CCs in the EN and WN are small enough that the
  nuclear components can be clearly differentiated in our efforts to
  estimate separations as small as 0.1\,mas. If this condition is not
  obeyed by the data then the CC clustering method cannot be used to
  measure the smallest separations, $\sim 0.1$\,mas. A priori we do
  not know if this assumption is consistent with the data. Ultimately,
  this assumption is verified empirically. For example, we present 7
  CC scatter plots of the nuclear region in Figures~4, 7, 8 and
  9. There is a strong tendency for the adjacent background to be
  almost devoid of any CCs above 20\,mJy in the surrounding regions as
  well as regions that exist between clusters of strong CCs $\leq
  0.1$\,mas across in the east west direction. The north-south
  resolution is worse making the constraint on the north-south
  distribution of CCs comprising the WN and EN more subjective in
  nature. The north-south distributions of CCs in Figures~4, 7, 8 and
  9 also seem to naturally cutoff at a size $ < 0.15$\,mas, similar to
  the resolution of the longest north-south baselines ($\sim 0.13 -
  0.14$\,mas).

The details of our CC models of the WN and EN can be found in
Table~2. The first two columns are the date of the observation and the
number of days since the nuclear separation was first detected. This
is not the same as the time from the epoch of physical separation due
to the finite resolution of the array. The next two columns are the
flux densities of the WN, EN and the total of these numbers,
respectively. The absolute flux calibration accuracy of the BU data is
about 5\% \citep{jor17}\footnote{Normally, the VLBA-BU-BLAZAR
    fluxes are accurate to within $\sim5\%$, but from early 2019 to
    Septamber 2019 it is quite a bit more uncertain, at least 30\%,
    because of pointing errors. Given the diffuse nature of 3C 84, we
    can only roughly estimate corrections through use the Mets\"ahovi monitoring program fluxes at 37\,GHz \citep[for a
      description of the Mets\"ahovi monitoring, see][]{ter04}. The
    VLBA resolves out too much of the structure to do that accurately,
    as we can for more compact sources.}. Column 5 provides
  the properties of the uniformly weighted beam on top and the
  naturally weighted beam below in parenthesis. Column (6) is the
peak intensity from the CLEAN images (Figure 2) for comparison with
the numbers in the previous columns. Columns (7) and (8) are the
largest magnitude negative contour values from the CLEAN image and the
CC cutoff used in or models, respectively. Columns (9) - (11) are the
uncertainties in the nuclear separation that are computed per the
methods of Section~3.2.

First, let us consider the epoch of August 26 2018. Since the absolute
astrometry is lost in the phase self-calibration step of the VLBI data
reduction, we are only interested in the relative positions of the EN
and WN and we place the brighter WN at the origin. We include
all the CCs within a region $\leq 0.1$\,mas in the east-west
direction. The blue (red) ellipse encircles all the CCs that are
associated with the EN (WN). The EN involves considerably weaker CCs
than the WN and the distribution tends to be more elongated in the
north-south direction. This is typical during the first year of our
monitoring of the time evolution of the nuclear region. The May 12
2019 observation in the top righthand side of Figure~4 shows a larger
separation between the EN and WN.

The bottom two panels of Figure~4 show the emergence of a third
component in the fall of 2019. It was clearly seen in the October 2019
observation, but the SNR was poor so we show the high SNR epochs of
November 2019 and January 2020. The bottom right hand panel shows that
the centroids of the three clusters of CCs (see the next subsection)
are nearly colinear. Thus, it makes sense to consider them all part of
the same flow. As such, we tentatively relabel the three components as
follows. Based on the early epochs, the WN is associated with the
point of origin. It is the brightest feature in 2018 and the much
weaker EN seems to flow out of the partially resolved cluster in the
time frame of August to December. We therefore consider the EN as the
first ejection, or ejection A. In the fall of 2019, the new feature
shown inside the green ellipse is about half the brightness of the EN
and WN, we call it ejection B. We actually never see it separate from
the WN, so this scenario is not verified, but suggested by the pattern
evolution. We cannot rule out that the red ellipse represents
a new ejection in the western direction and the green component is the
point of origin. This is disfavored because it would require the EN to
be contracting towards the point of origin and the point of origin
would change from being a much brighter component than the EN to much
fainter than the EN.

\begin{figure}
\begin{center}
\includegraphics[width=87 mm, angle= 0]{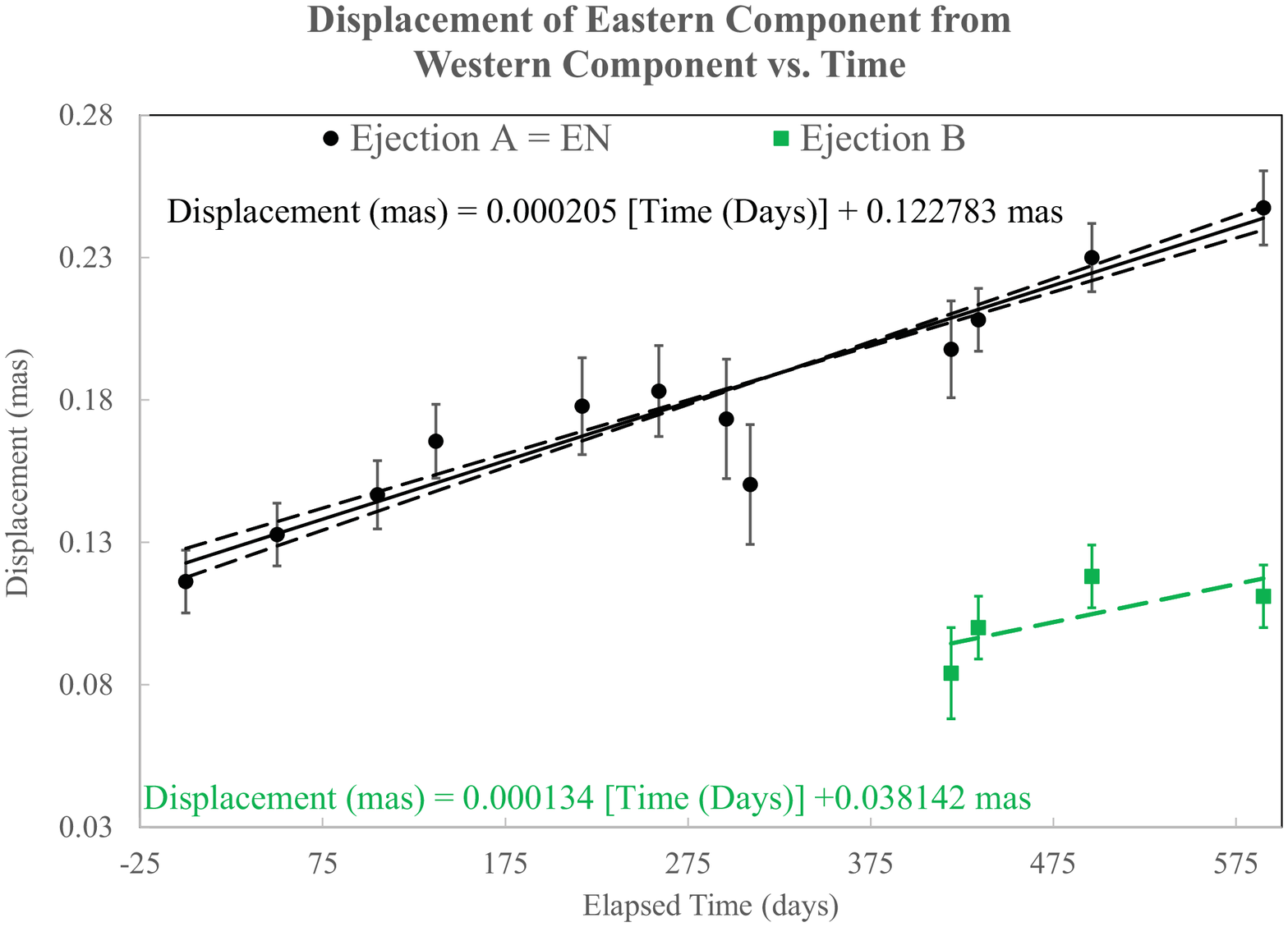}
\includegraphics[width=87 mm, angle= 0]{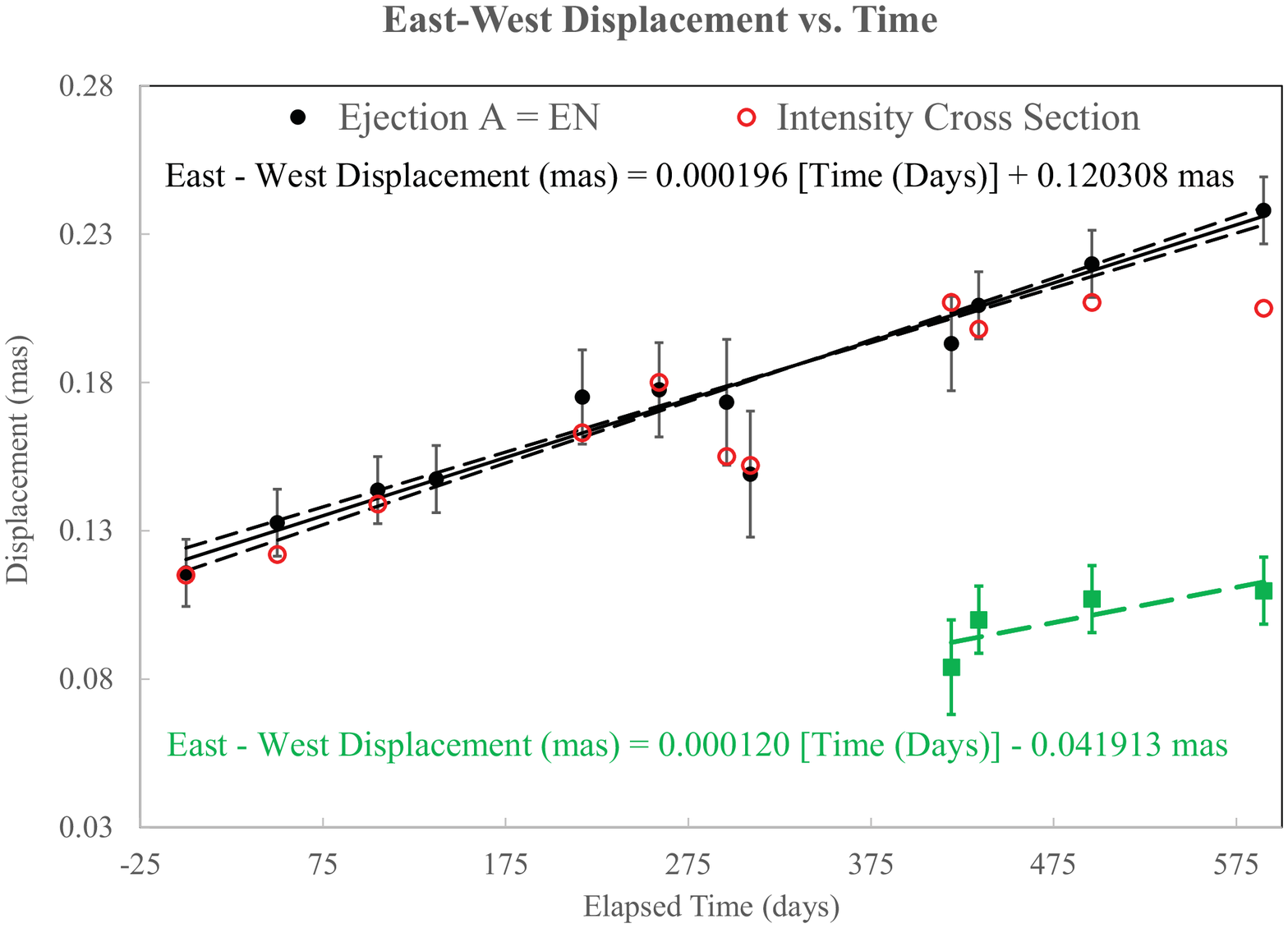}
\includegraphics[width=87 mm, angle= 0]{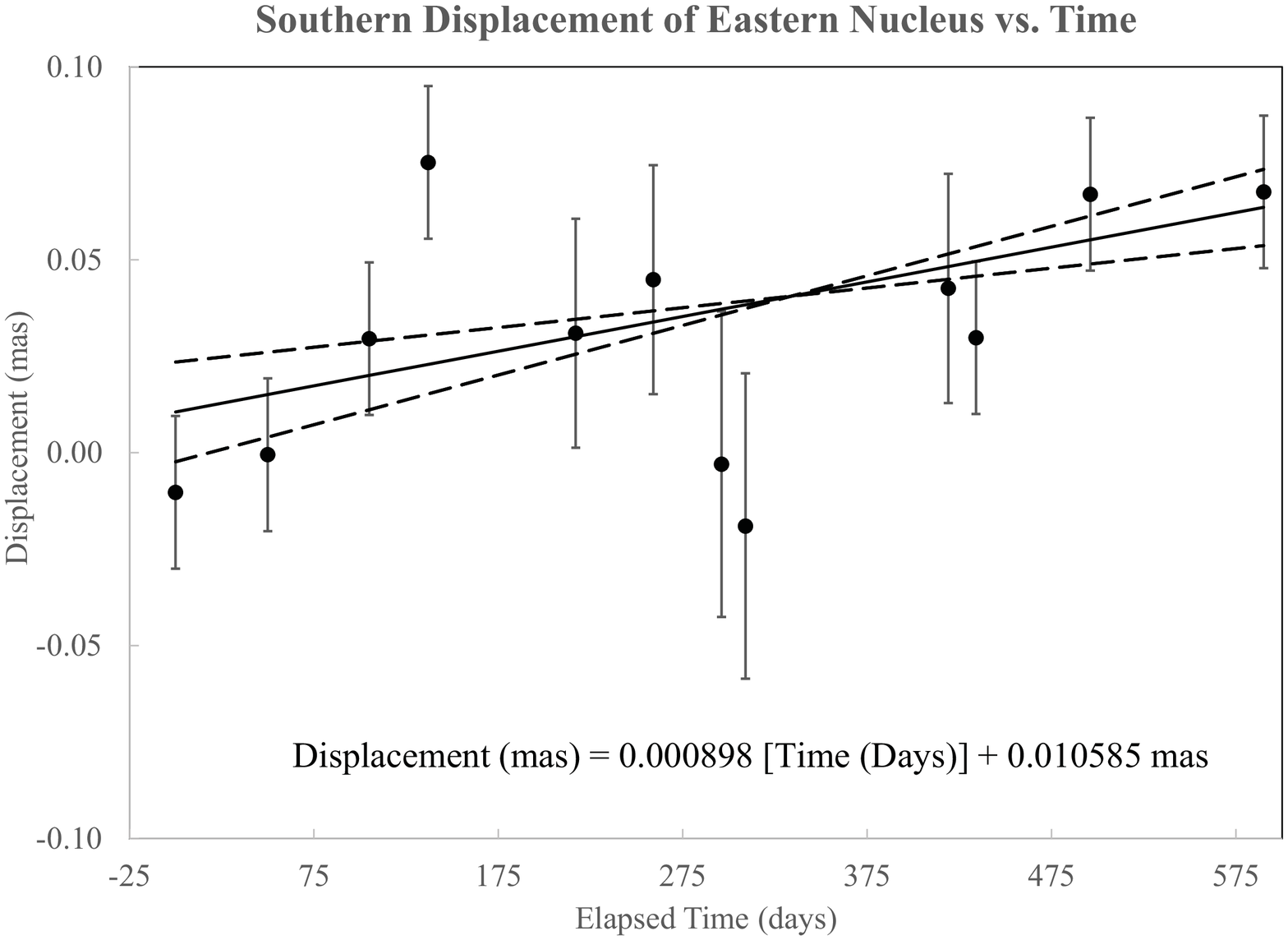}
\caption{\footnotesize{The separation between WN and EN versus time is
    demonstrated in two ways. The top left panel is a plot of
    total displacement versus time. The east-west displacement is
    plotted in the top right panel. The two methods yield a
    similar separation velocity. The solid black line is a least
    squares fit with uncertainty in the vertical variable
    \citep{ree89}.  The dashed lines define the standard error of the
    fit. The equation for the best fit is printed on the plot. We also
    provide a crude fit for ejection B in green. The top
      right panel also plots the east-west separation from the
    intensity profiles (in red) with no information from January 10,
    2019 available from Figure~3. We consider these data less rigorous
    and are shown as a consistency check. The peak separation of the
    intensity cross-sections lies below the uncertainty of the CC
    model fit for April 2020 (day 590). This is a consequence of a
    curved trajectory of the EN. As can be seen in Figure 2, the EN
    starts veering south relative to the declination of the WN in 2020
    and is most pronounced in April. Thus, there is no east-west cross
    section that captures intensity peaks of both the EN and WN. The
    April data indicates the breakdown of the simple east-west cross
    section method for a curved trajectory for the
    ejection. For completeness, we include the southern
      displacement of the EN with its fit in the bottom panel. The
      magnitude of the trend is small compared to the uncertainties
      and this is reflected in the standard error of the best
      fit. Thus, the result should be viewed with caution. The three
      outliers are January 10, 2019 (that is discussed in Appendix B)
      and the two low SNR observations in June and July 2019.}}
\end{center}
\end{figure}

\subsection{The Position of the Features}

We calculate the position of the features by taking the flux-density
weighted mean (centroid) of the CC positions in the feature. In Figure
4, the red triangle inside the red ellipse and the blue triangle inside the blue ellipse are the locations of the
centroids of the flux density of the CCs encapsulated within each
ellipse. We assign an uncertainty to the centroid location. The origin of the coordinates will be the centroid of the WN and all displacements are measured relative to this. The uncertainty in the positional locations is $\sim 10\%$ of the
synthesized beam FWHM for these bright components
\citep{lis09} \footnote{see also
  https://science.nrao.edu/facilities/vla/docs/manuals/oss/performance/positional-accuracy}. In
Table 2, the uncertainties in coordinate separation, $\sigma_{\Delta
  X}$ and $\sigma_{\Delta Y}$ ($X$ is right ascension and $Y$ is
declination), are the uncertainties on the individual centroid
coordinates added in quadrature. Similarly, the uncertainty of the
total displacement between the EN and WN, $\sigma_{D}$, is calculated
as the error propagation of the centroid uncertainties. The random noise contribution to the separation uncertainty can be
  estimated by $\sigma_{D} (\rm{noise})\approx$ 0.5 d
  $\sqrt{\sigma_{\rm{rms}}/S_{\rm{peak}}}$, where $\sigma_{\rm{rms}}$
  is the post-fit rms noise of the image and $S_{\rm{peak}}$ is the
  flux density of a putative unresolved component of size $d$
  \citep{fom99,lee08}. $\sigma_{\rm{rms}}<$ 1 mJy for all epochs
  except June and July 2019 where it is closer to 2 mJy. We determine
  in Table 2 that $S_{\rm{peak}}> 479$ mJy so all the SNR are $> 500$
  except June 2019 which is $\gtrsim 400$. It is therefore clear
    that positional uncertainty is not dominated by the random noise
    component. The error estimate adopted in Table~2 is much more conservative than
  stochastic noise errors and incorporates systematic uncertainty.

\section{The Separation Rate of the Double Nucleus}

The motivation for defining the locations of the EN and WN and the
corresponding uncertainty is to track the time evolution of the
separation of the components. This is performed in Figure~5. The top left panel is the plot of the displacement versus
time. However, we note that the displacement is primarily east-west
and the highest resolution is in this direction. We plot east-west (x
direction) displacement in the top right panel for direct
comparison to the intensity cross-sections in Figure~3.

The data were fit by a least squares fit with uncertainty in the
vertical variable \citep{ree89}. The solid line is the fit and the
dashed lines represent the standard error to the fit. Based on the fit
and standard error, the time averaged nuclear separation speed is
$0.086 \pm 0.008$\,c. The fit to the total displacement yields a
separation rate of $73 \pm 7$\,$\mu\rm{as}$ per year compared to $70
\pm 6$\,$\mu\rm{as}$ per year for pure east-west displacement. Note
that we also plot (in green) the separation of the second ejection,
ejection B, in the lower right hand corner of the two panels. We also
provide very crude four point fits to the ejection velocity.

It is apparent that the EN-WN separations in the spring and summer of
2019 (days 217 $-$ 309) deviate from the least squares fits in
Figure~5. This is not unexpected from the Tables~1 and 2. These are
the most compromised observations. However, the July 2019 estimated
separation lies particularly far from the fit. It is possible to
explain this behaviour to be due to the ejection``B" that is shown in
the bottom frames of Figure~4. This central component is resolved from
EN and WN in our CC cluster models first in October 2019, but it is
most likely present already in earlier epochs. If this is indeed the
case than we may just not resolve it in July 2019, but it is there and
it blends with WN, moving its measured position slightly towards
EN. This can cause the small inward motion seen at that epoch. The
results of this paper do not rely on this interpretation.

By considering the east-west displacement, we are able to directly
compare the results of the intensity profile description in Figure~3
with the CC models of the EN and WN in Section~3. The
  distribution of intensity cross-section separations is generally
  consistent with both the standard error of the least squares fit and
  the uncertainty of the CC model fits to the EN-WN separation (the
  measurement uncertainty given in Table~2) except for two regions, in
  June and July 2019 for which the SNR was poor and the last two
  points, January and April 2020. Figure~2 and the bottom frame of
  Figure 5 seem to indicate that beginning in January 2020, the EN no
  longer appears to follow a predominantly east-west trajectory, but
  begins to veer southward towards the southerly directed large scale
  jet. The EN appears to propagate in a southeasterly direction
rather than an easterly direction. Thus, a pure east-west cross
section is unable to properly capture the intensity peaks of both
components. The situation is more extreme in April and this could
cause the measurement using the intensity cross section to be less
than the true east-west displacement. This circumstance illustrates
the limitation of the simple east-west cross section method of
measuring separation if the trajectory curves.

 For the sake of completeness, we plot the southern
  displacement of the EN as it separates from the WN in the bottom
  panel of Figure~5. The putative trend is far less pronounced than
  the east-west separation and the uncertainty is much larger (the
  synthesized beam is elongated in this direction). Consequently,
  there is no robust quantitative assessment that can be made due to
  the limitations of the CC based method of this paper.

One can also perform a long extrapolation back in time
  (assuming that the separation velocity was approximately constant in
  time) of the linear fits in order to find the epoch of zero
  separation from the nucleus. From Figure~5, we have that ejection A,
  the WN, emerged on January 1, 2017 with a large uncertainty that
  places it anywhere between September 21, 2016 and March 22, 2017. We
  can crudely estimate the zero separation epoch for ejection B from
  Figure~5 as $\sim$\,November 6, 2017. The two ejections originated at
  least 7 months apart and represent two distinct nuclear events. Our
  estimate of the January 1, 2017 ejection is not tightly constrained,
  but we would be remiss not to mention that there was a major gamma
  ray flare reaching TeV energies that was detected by the Large Area Telescope on board the Fermi satellite
  and MAGIC telescopes that peaked around December 31, 2016 to January
  1, 2017 \citep{bag17,ans18}.

\section{The Double Nucleus in the uv-Plane}

It is customary for astronomers to fit the visibility data
    with a small number of Gaussians in the order to track component
  separations. Thus, we explore the possibility of evidence of the
double nucleus in the $(u,v)$ plane by making Gaussian fits to the
nuclear region. However, there are two concerns for this method in the
context of the separating nuclear features in
Figure~2. Firstly, the radio source 3C 84 is extremely complex
  compared to a typical blazar nucleus and it is extended. So,
one needs to use a CC model to represent it, but that creates a
problem regarding how large of an area around the core should be
cleared from CCs before model-fitting. This is always a bit subjective and can influence the results
significantly. The second issue is that the initial stages of
  the expansion in Figure~2 are at the limits of the resolution of the
  VLBA.  There is no unique solution to the fitting
method. Each ambiguity and assumption adds a possible error to
  the fitting process. After some initial attempts to make Gaussian
  fits to the nuclear region we found two methods that give reasonable
  results.

\subsection{Gaussian Fitting Method 1}
The first method addresses the issue of the complex background
  emission by a simple excision method that can be uniformly applied
  to each epoch. We took the CLEAN model and automatically removed CCs
  from the area three times the uniformly weighted beam size (from
  Table~2) around the core. The excised region is then modelled with
  circular Gaussians. We added new components in the model if a
    clear peak was seen in the residual image after fitting. The
  background subtraction is not perfect so it introduces additional
  Gaussian components in addition to those associated with the EN, WN
  and ejection B. This method typically needs many components to
    represent the core, which can make the fit unstable.

\begin{deluxetable*}{cccccccccc}
\tablecaption{Gaussian Fits to the Double Nucleus}
\tabletypesize{\tiny} \tablehead{ \colhead{(1)} & \colhead{(2)} &
  \colhead{(3)} & \colhead{(4)} & \colhead{(5)} & \colhead{(6)} &
  \colhead{(7)} & \colhead{(8)} & \colhead{(9)} \\ \colhead{Date} &
  \colhead{Model} & \colhead{$F_{\nu, \mathrm{WN}}$} &
  \colhead{$F_{\nu,\mathrm{EN}}$} & \colhead{$\Delta \rm{RA}$} &
  \colhead{$\Delta \rm{dec}$} & \colhead{Separation} &
  \colhead{Gaussian FWHM WN} & \colhead{Gaussian FWHM EN}
  \\ \colhead{} & \colhead{} & \colhead{(Jy)} & \colhead{(Jy)} &
  \colhead{(mas)} & \colhead{(mas)} & \colhead{(mas)} &
  \colhead{(mas)} & \colhead{(mas)} }
\startdata
08/26/2018 & 1 & 1.137 & 1.437 & 0.093 & 0.046 & 0.104 & 0.073 & 0.156 \\
08/26/2018 & 2 & 1.146 & 0.840 & 0.135 & 0.001 & 0.135 & 0.068 & 0.102 \\
10/15/2018 & 1 & 2.525 & 1.138 & 0.157 & -0.053 & 0.166 & 0.104 & 0.127 \\
10/15/2018 & 2 & 2.219 & 1.459 & 0.146 & -0.008 & 0.146 & 0.088 & 0.134 \\
12/09/2018 & 1 & 3.211 & 1.435 & 0.166 & -0.027 & 0.168 & 0.113 & 0.052 \\
12/09/2018 & 2 & 2.819 & 1.856 & 0.151 & -0.034 & 0.155 & 0.085 & 0.082 \\
01/10/2019 & 1 & 1.179 & 4.634 & 0.139 & 0.019 & 0.140 & 0.085 & 0.190 \\
01/10/2019 & 2 & 2.030 & 2.321 & 0.127 & 0.018 & 0.128 & 0.076 & 0.072 \\
03/31/2019 & 1 & 1.193 & 1.576 & 0.169 & -0.009 & 0.169 & 0.068 & 0.132 \\
03/31/2019 & 2 & 1.819 & 0.932 & 0.166 & 0.022 & 0.167 & 0.066 & 0.071 \\
05/12/2019 & 1 & 2.735 & 0.593 & 0.200 & -0.007 & 0.200 & 0.143 & 0.018 \\
05/12/2019 & 2 & 2.601 & 0.786 & 0.199 & -0.043 & 0.204 & 0.119 & point source \\
06/18/2019 & 1 & 2.794 & 0.970 & 0.170 & -0.010 & 0.170 & 0.091 & 0.011 \\
06/18/2019 & 2 & 2.470 & 0.795 & 0.151 & 0.018 & 0.152 & 0.074 & 0.107 \\
07/01/2019 & 1 & 2.716 & 1.961 & 0.162 & 0.007 & 0.162 & 0.019 & 0.055 \\
07/01/2019 & 2 & 3.876 & 2.032 & 0.151 & 0.028 & 0.153 & 0.071 & 0.029 \\
10/19/2019 & 1 & 2.505 & 1.629 & 0.183 & -0.066 & 0.195 & 0.155 & 0.100 \\
10/19/2019 & 2 & 1.284 & 1.454 & 0.221 & -0.026 & 0.223 & 0.091 & 0.092 \\
11/03/2019 & 1 & 0.688 & 1.041 & 0.234 & -0.040 & 0.237 & 0.063 & 0.053 \\
11/03/2019 & 2 & 1.193 & 1.120 & 0.222 & -0.031 & 0.224 & 0.061 & 0.042 \\
01/04/2020 & 1 & 1.370 & 1.109 & 0.232 & -0.066 & 0.242 & 0.104 & 0.059 \\
01/04/2020 & 2 & 1.222 & 1.523 & 0.240 & -0.064 & 0.249 & 0.075 & 0.081 \\
04/07/2020 & 1 & 2.517 & 0.706 & 0.264 & -0.111 & 0.286 & 0.144 & point source \\
04/07/2020 & 2 & 1.878 & 1.286 & 0.262 & -0.080 & 0.274 & 0.109 & 0.046
\enddata
\end{deluxetable*}

\subsection{Gaussian Fitting Method 2}
We took the CLEAN model, but removed only those CCs that
  roughly correspond to the double (triple) nucleus in a
  super-resolved image (created with a 0.1\,mas circular beam). This
  CC excision was done ``by eye", so there is some subjectivity
  here. We modelled this region with 2 or 3 circular Gaussian
  components. The method is less subjective than the first as a result
  of incorporating additional information from the super-resolved
  image. However, the method relies significantly on the CLEAN model,
  and it is therefore not independent.

\begin{figure}[htp!]
\begin{center}
\includegraphics[width=120 mm, angle= 0]{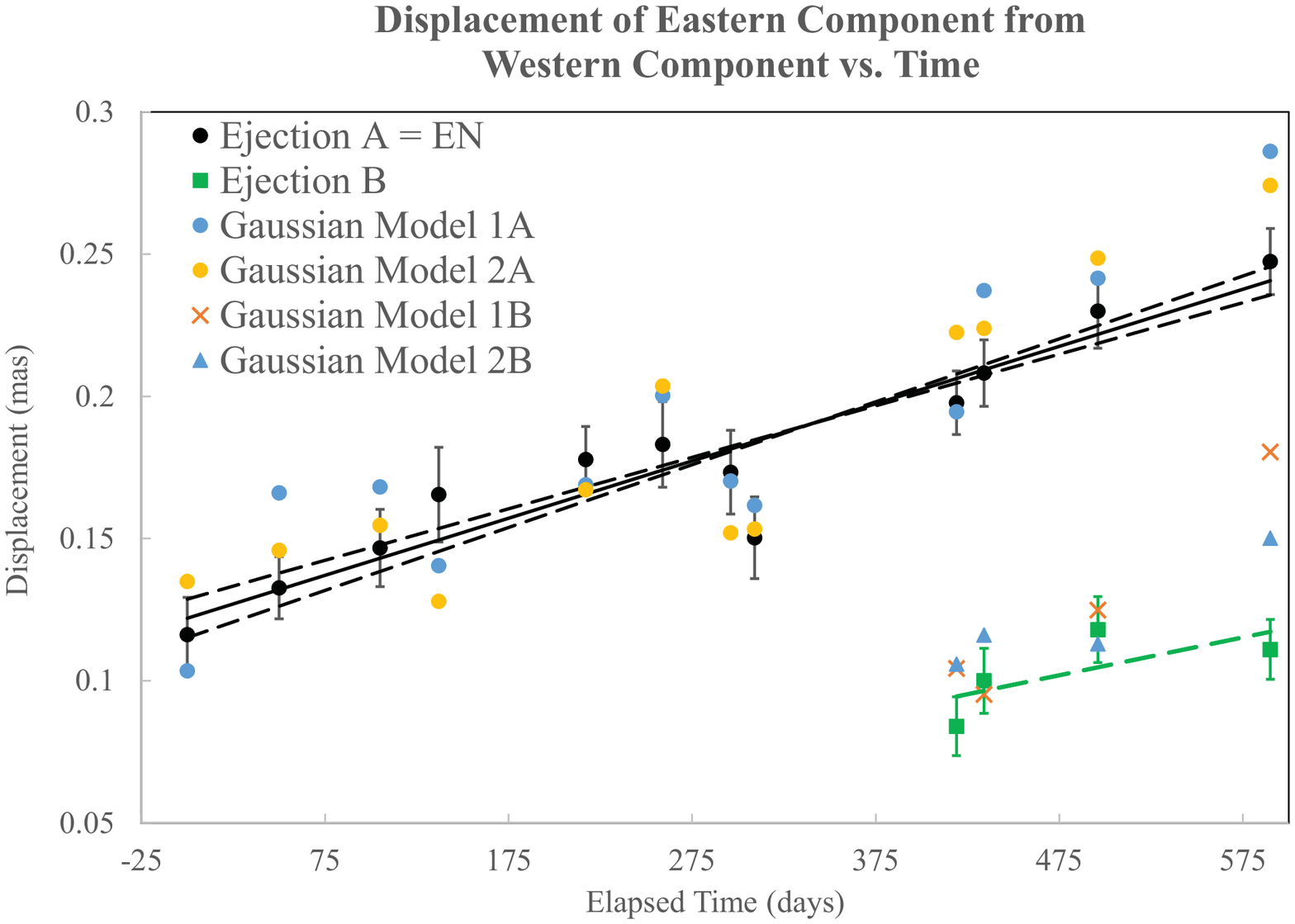}
\includegraphics[width=120 mm, angle= 0]{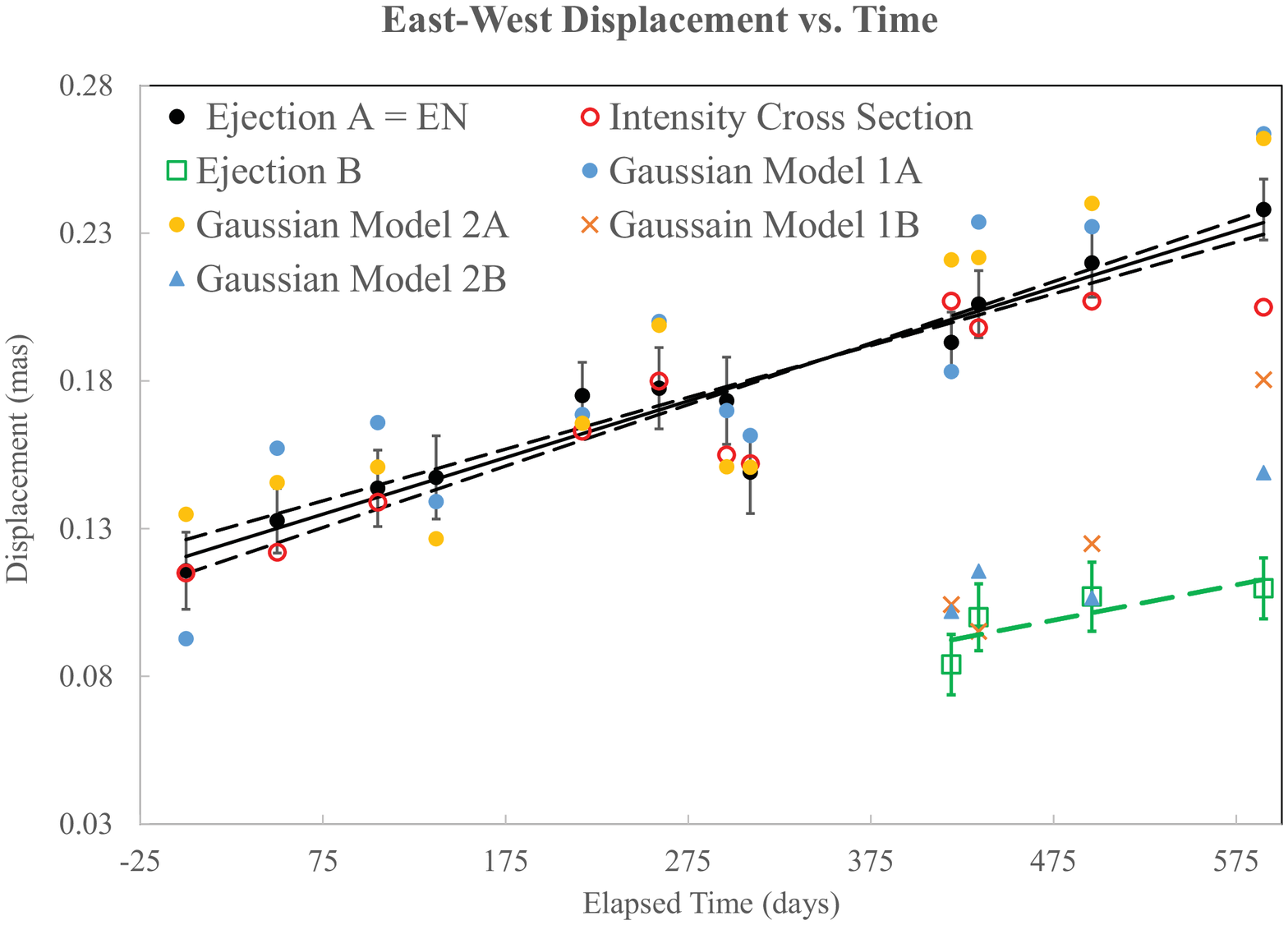}
\footnotesize{\caption{The Gaussian components produced by the Difmap
    fit in the visibility domain are indicated by blue and orange
    symbols, for Method 1 (described in Section~5.1) and Method 2
    (described in Section~5.2). These are overlayed on the plots from
    Figure~5. Method 1A and 2A points are the distances from the
    Gaussian representing the EN from the Gaussian representing the
    WN. Similarly, method 1B and 2B points are the distances from the
    Gaussian representing the ejection B from the Gaussian
    representing the WN. We do not present error bars on the Gaussian
    fit data since it would clutter the graph making it
    unintelligible.}}
\end{center}
\end{figure}

\subsection{Comparison of Methods}
In Figure~6 we reproduce Figure~5 with the addition of the
  separation derived from the Gaussian models of the nuclear region
  described above. There are two things to note. First of all, the
  east-west separation in 2018 and early 2019 displayed in the bottom
  panel deviates much more from the cross-sectional method of Figure 3
  than the CC approach used in this paper. Secondly, because Model 2
  is derived using super-resolved images, it agrees better with the CC
  model data. The Gaussian fits to the EN and WN are listed in Table
  3. The two fits can be compared to each other as well as the CC
  based method in Table 2. There are some large differences in the
  flux density of the EN and WN between the two models in columns (3)
  and (4). This might be a consequence of the different methods of
  removing the background and the relatively large Gaussian FWHM in
  columns (8) and (9) that are comparable to the separations in column
  (7). If we perform a linear fit to the separation from the Gaussian
  fit we find a similar separation rate for both Gaussian fitting
  schemes, about 15\% larger than what was found in Section 4,
  $0.100$\,c. This is surprising considering that on a point by point
  level the agreement is not that tight.

In Figure~7, the Gaussian components for both models given in
  Table~3 are overlayed on CC distribution for two adjacent epochs. We
  define EN and WN as in Figure~4. The image is qualitative, but it
  gives the reader a feel for the difference between the two model
  results and the difference between the models and the CC based
  methods. Method 2, which utilizes information from the
  super-resolved images, seems very consistent with the EN and WN
  identifications in May 2019. Figure~6 and Tables~2 and 3 indicate
  some significant discrepancies between the results of the CC method
  of Sections 3 and 4 and the Gaussian models during the epoch April
  7, 2020. This is most pronounced for ejection B. In Figure~8, we
  overlay the Gaussian fits on the CC scatter plot from
  Figure~4. There is clearly disagreement between the two
  fits. Furthermore, Model 1 seems to blend the EN and ejection B and
  Model 2 seems to ignore all the CCs at the north end of the EN. This
  suggests that the Gaussian fit methods have difficulty resolving 3
  components within a total extent of 0.24\,mas. The figure clearly
  shows the origin of the larger displacements for ejection B in the
  April 2020 epoch of Figure~6 given by the Gaussian models compared
  to the CC based analysis.

The main objective of this study is to track the time
  evolution of the double separation as close to its inception as
  possible in order to estimate the trajectory and separation speed as
  accurately as possible. There is arbitrariness in the Gaussian
  fitting procedure that is accentuated in the early epochs (in 2018)
  when the component separation was smaller. The intensity
  cross-sections were introduced in Figure~3 to provide a clear
  diagnostic of the model fitting schemes. Based on Figure~6, both
  Gaussian decomposition schemes give worse fits to the east-west
  separation derived from the intensity cross-sections in the early
  epochs than the CC based methods of Section 3 and 4. We do not
  advocate extrapolating the Gaussian fitting method to the first
  three epochs of small component separation. Similarly, Figure~8
  seems to indicate that the second ejection crowds the narrow
  0.24\,mas field making it difficult for the Gaussian fits to
  segregate components uniquely or with high accuracy. For this
  reason, we consider the CC method of this paper preferable to
  conventional Gaussian fitting for exploring the partially resolved
  compact nuclear region of 3C\,84. The Gaussian fitting does provide
  qualitative agreement with CC model fitting. As such, along with the
  intensity cross-sections, this corroborates the results of the CC
  based analysis. We have also demonstrated that with proper care in
  the definition of the Gaussian models, one can achieve a similar
  estimate (within 15\%) to the nuclear separation.

\begin{figure}
\begin{center}
\includegraphics[width=85 mm, angle= 0]{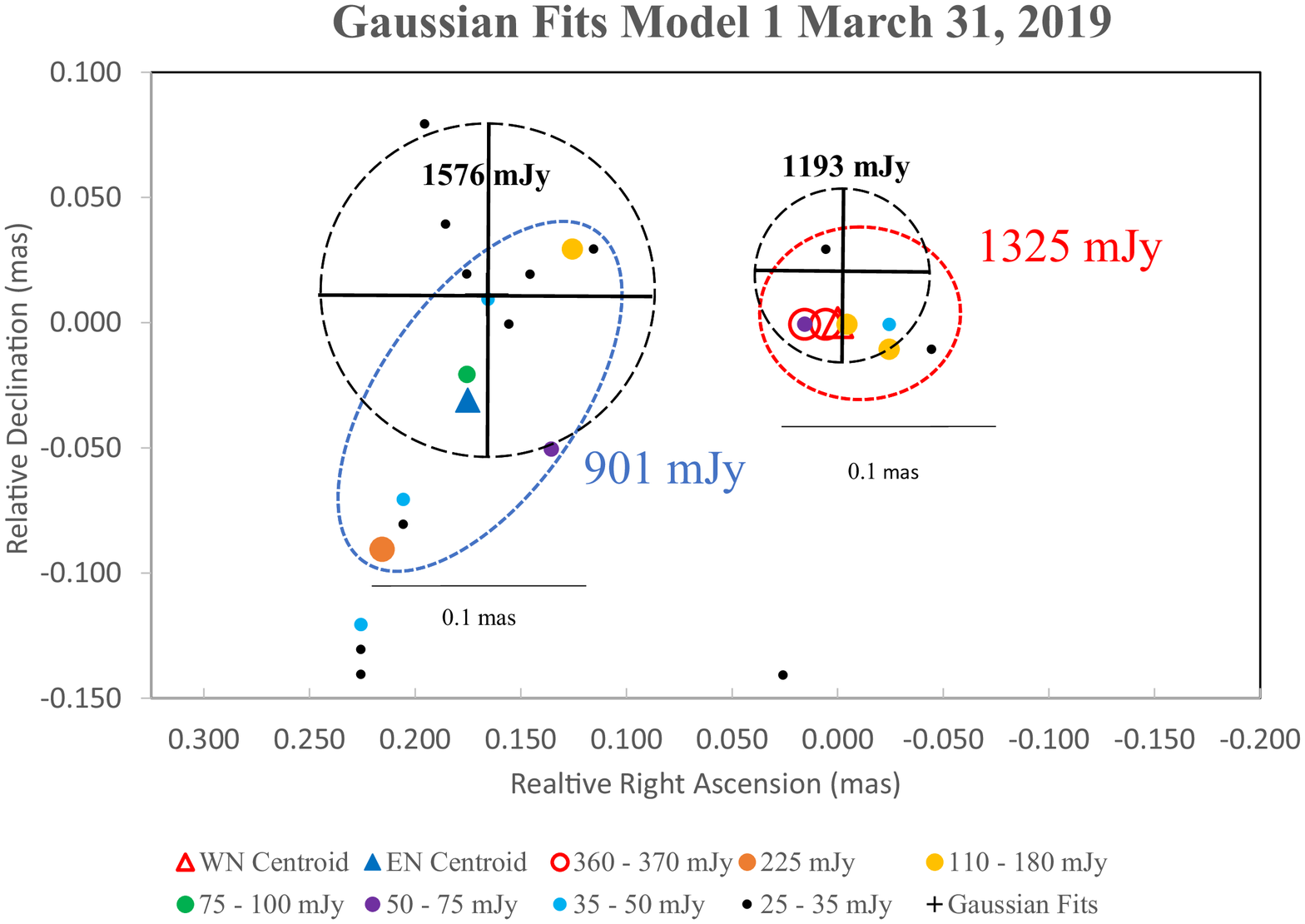}
\includegraphics[width=85 mm, angle= 0]{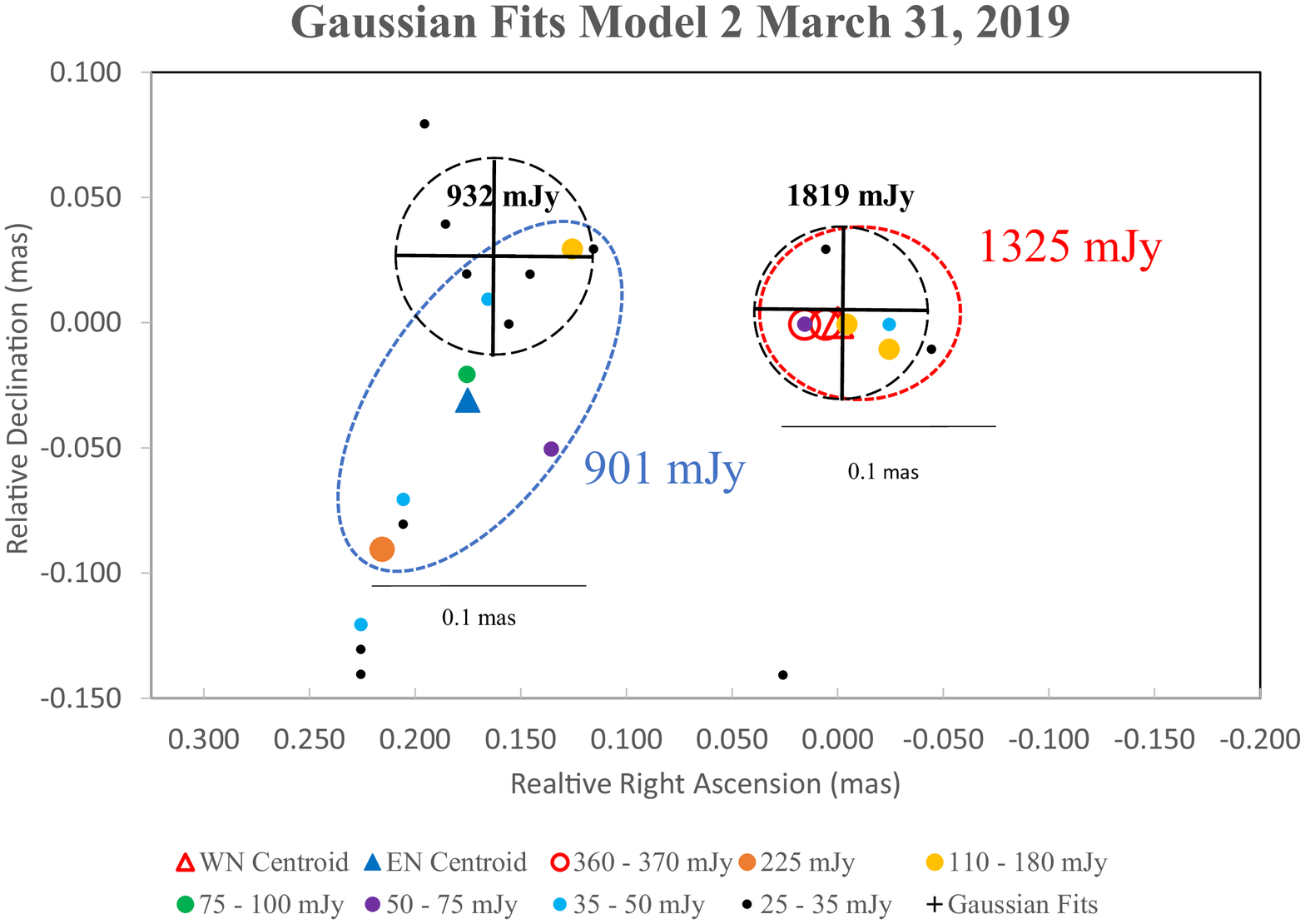}
\includegraphics[width=85 mm, angle= 0]{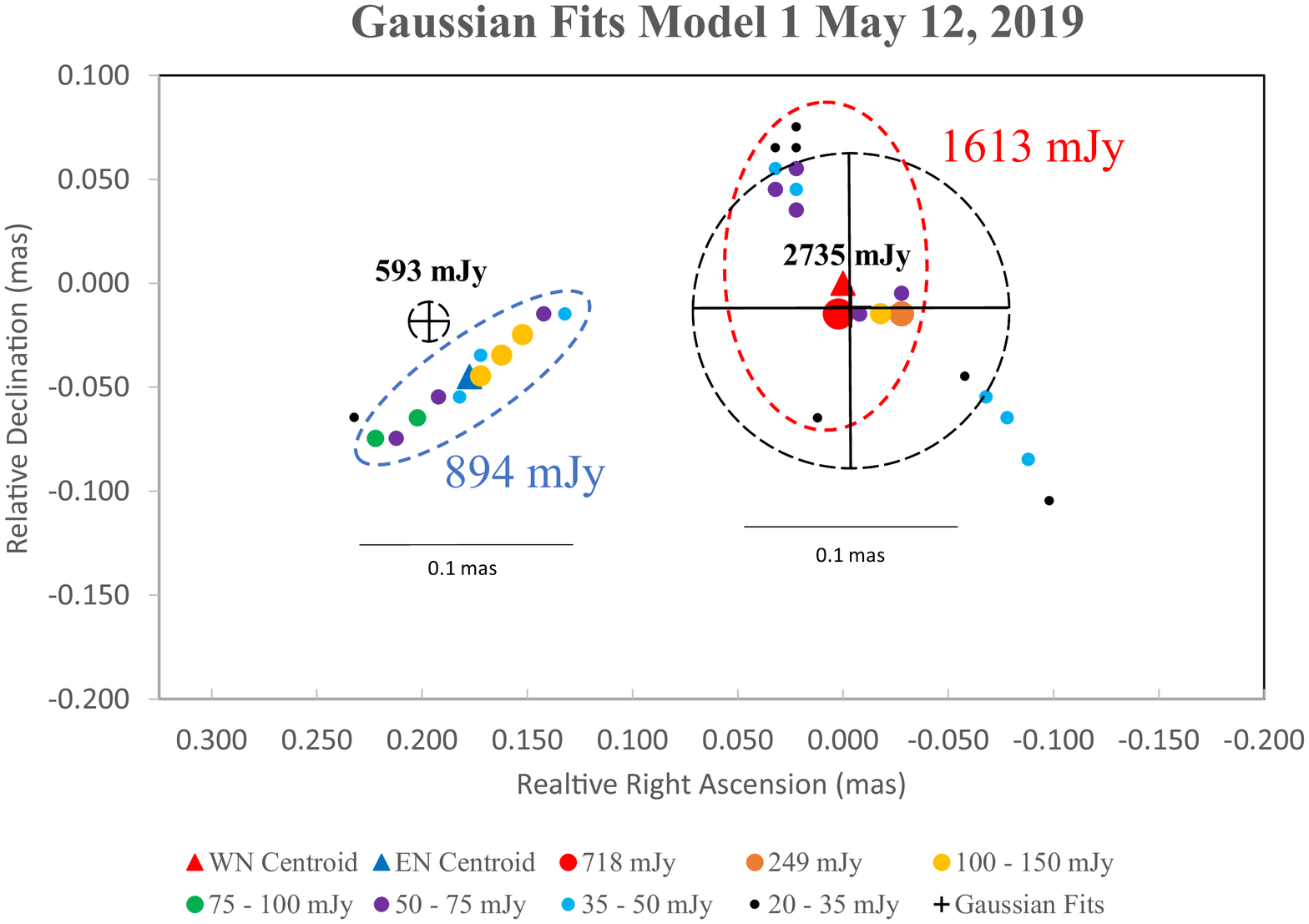}
\includegraphics[width=85 mm, angle= 0]{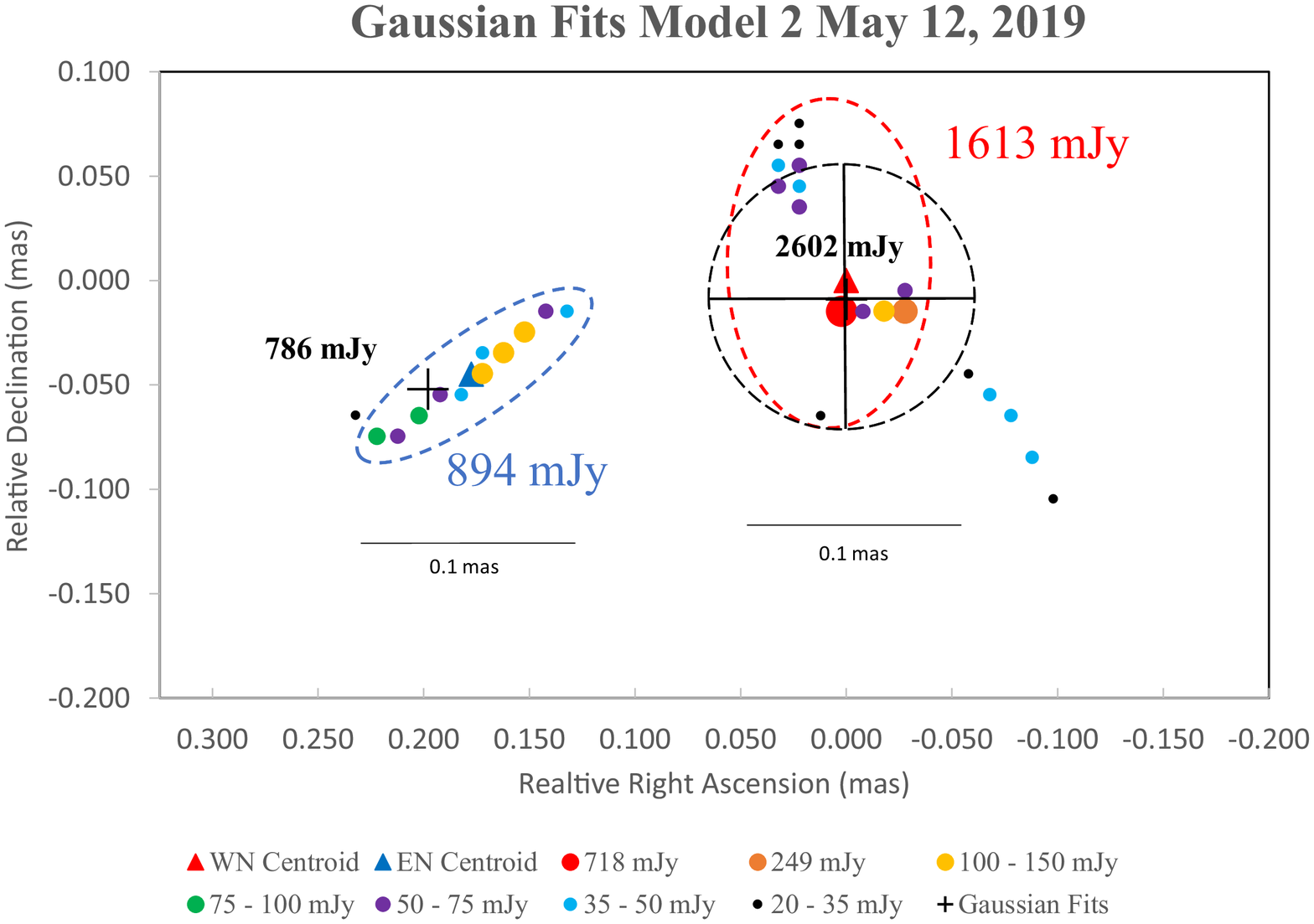}
\footnotesize{\caption{The foreground Gaussian components from
      Table~3 are indicated by the dashed black circles with the
      inscribed black crosses. Both Model 1 (described in Section 5.1)
      and Model 2 (described in Section 5.2) are plotted, but
      separately for clarity. The diameter of the circle represent the
      FWHM of the Gaussian component. These are superimposed on the CC
      component models as defined in Figure~4. The flux densities are
      in black near or inside each Gaussian component. The eastern
      component of Gaussian Model 2, in May, is a point source and it
      is represented by a black cross.}}
\end{center}
\end{figure}

\begin{figure}
\begin{center}
\includegraphics[width=85 mm, angle= 0]{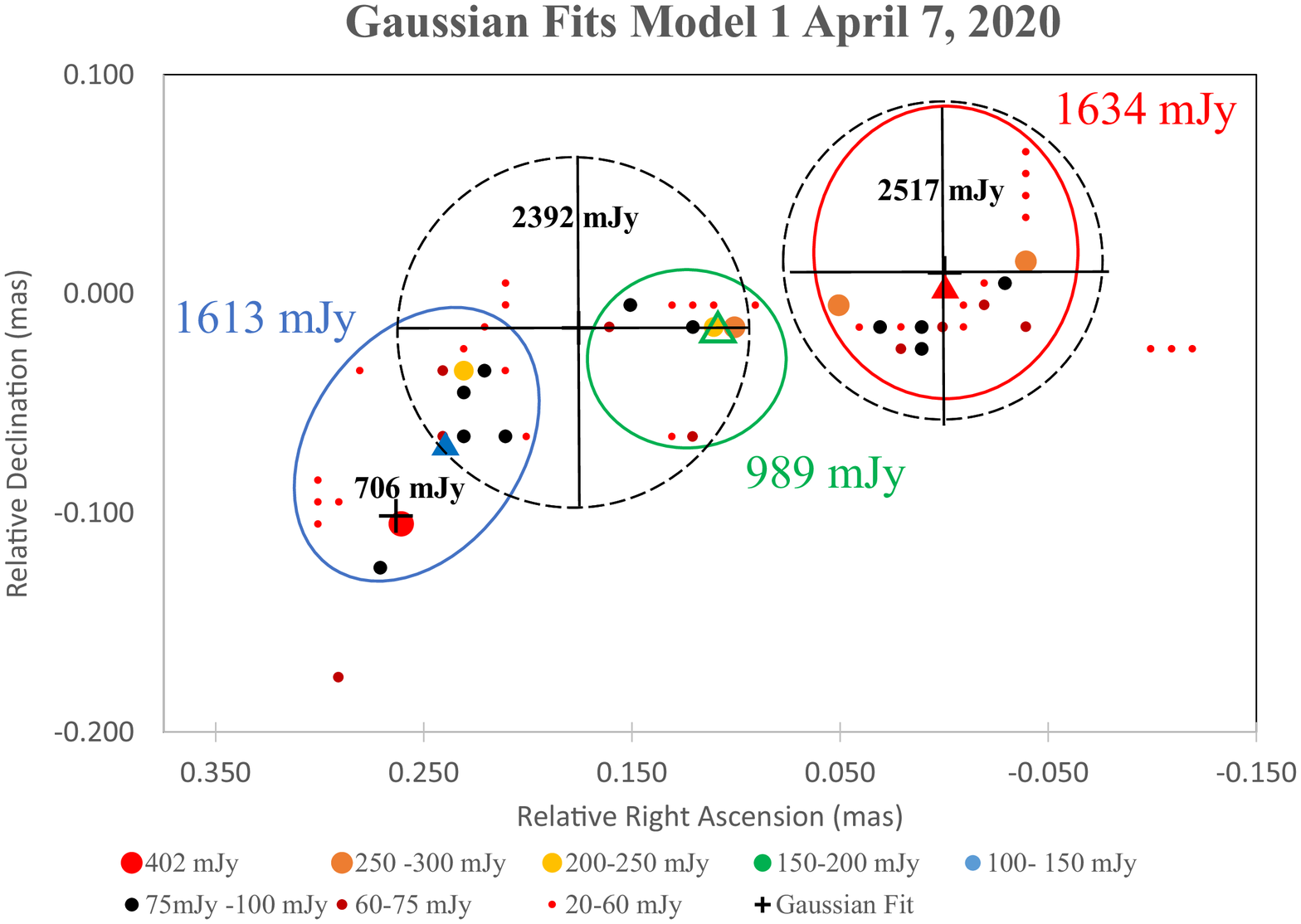}
\includegraphics[width=85 mm, angle= 0]{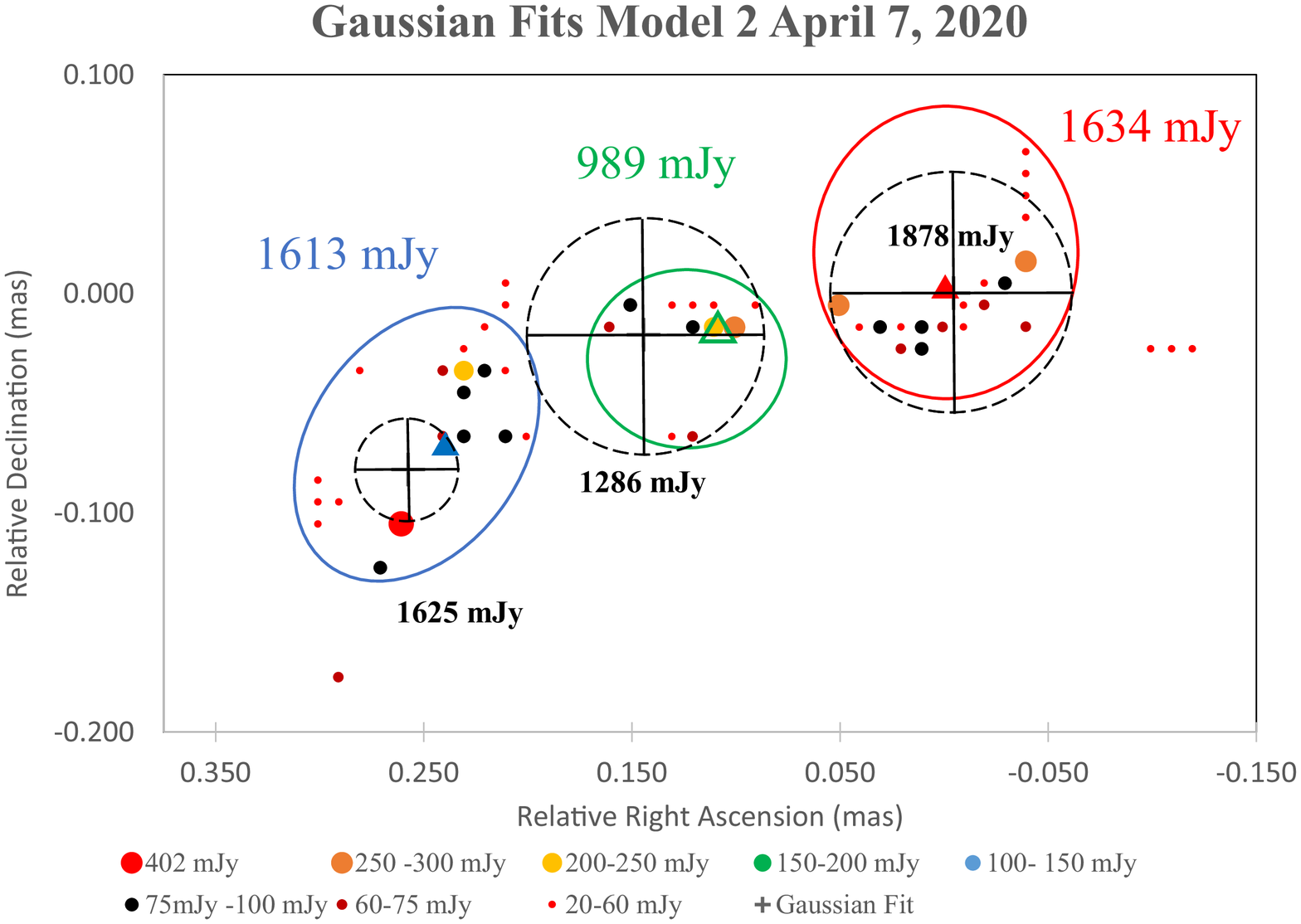}
\footnotesize{\caption{The figure compares the Gaussian models
      with the CC derived components during the epoch April 7,
      2020. The figure is formatted the same as Figures~4 and 7. The
      three bright components in close proximity to each other seems
      to provide a challenge for the Gaussian models as evidenced by
      the significant differences between the models from methods 1
      and 2. The eastern component of Gaussian Model 1 is a point
      source and it is represented by a black cross. We have not
      included the 0.1\,mas rulers in this figure in order to avoid
      distracting clutter.}}
\end{center}
\end{figure}

\section{The Nuclear Separation in the Context of RadioAstron}
In order to create more context for these results in 2018-2020, we
compare the nuclear structure to the triple nucleus observed on
September 21-22, 2013 by RadioAstron \citep{kar13,kar17}. First of
all, the RadioAstron observations were accompanied by a 43\,GHz
observation with an array of the VLBA combined with the phased VLA
\citep[][Savolainen et al., in prep.]{gio18}. We use the clean
component model from imaging of these data together with the methods
of this paper to detect evidence of a triple nucleus using the
east-west resolution of the VLBA. There were no September or October
observations by the BU monitoring program. The top panel of Figure~9
is a nuclear clustering model similar to Figure~4. It shows all the
CCs $>20$ mJy in a nuclear region that is a square, 0.4\,mas on a
side. Without the RadioAstron results, this compact configuration
would have been more ambiguous to decompose into three
components. Based on the RadioAstron images, we cluster the CC
components into three regions of concentration, RA East (blue), RA
Central (red) and RA West (green). The nuclear region is very clean,
there are relatively few CCs compared to the other epochs considered
and the background CC field is sparse near the triple nucleus in the
innermost 0.2 mas. This fortuitous circumstance facilitates the
identification of the components of the triple nucleus. The distance
from RA East to RA Central is $\approx 0.08$ mas and RA Central to RA
West is $\approx 0.09$ mas.

\begin{figure}
\begin{center}
\includegraphics[width=125mm, angle=0]{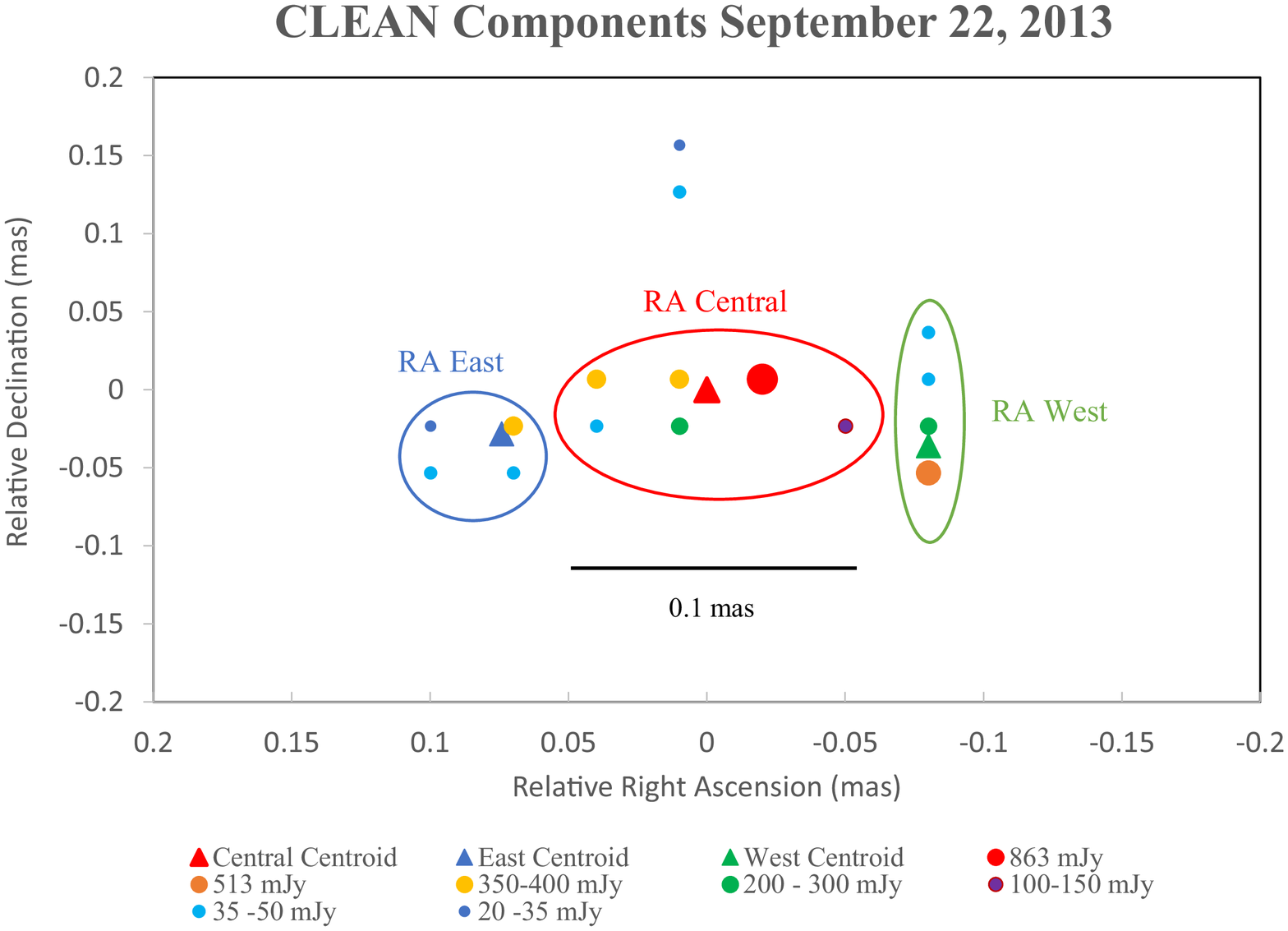}
\includegraphics[trim={5.3cm 0 0 0}, clip, width=120mm, angle=0]{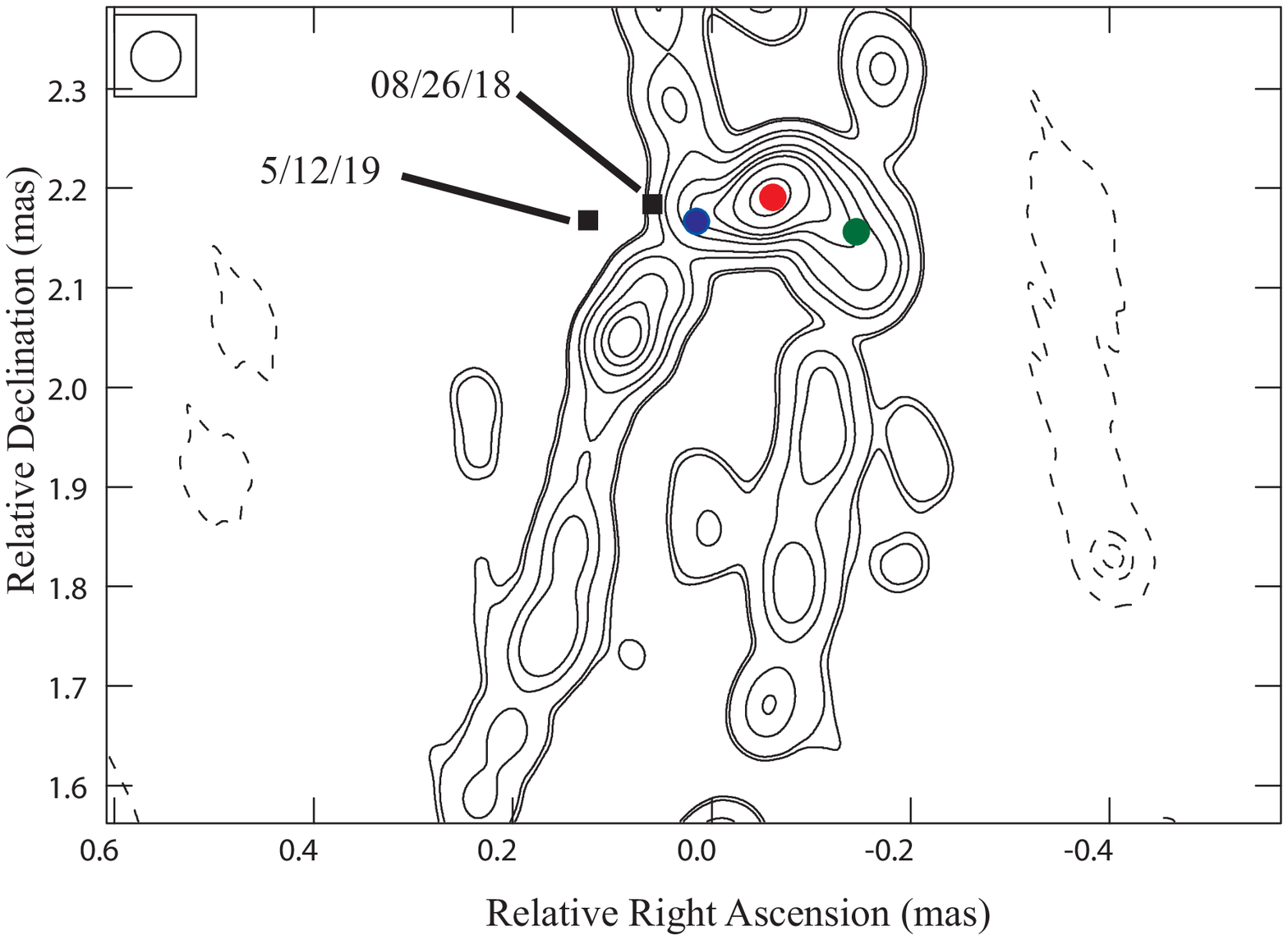}
\caption{The top panel shows the location of the nuclear CCs
  $>20$\,mJy on September 22, 2013 based on a 43\,GHz VLBA plus phased
  VLA observation that was quasi-simultaneous with the 22\,GHz RadioAstron observation in the bottom panel. We cluster
  the CCs as three physical components RA East (blue), RA Central
  (red) and RA West (green). Their positions are overlaid on the
  RadioAstron image from \citet{gio18}. We cross-identify the brightest
  component, RA Central, with the bright component in 2018/2019, the
  WN. Based on this identification, we also show the positions of the
  EN in August 26, 2018 and May 12, 2019 relative to the RadioAstron
  image in the bottom panel. The contour levels are (-10, -7.50, -3, 7.500, 10, 30, 50, 100, 150, 200, 300, 500)mJy.}
\end{center}
\end{figure}

We overlay the locations of the RA East, RA Central and RA West as
blue, red and green triangles, respectively from the top
panel on the RadioAstron image. The locations appear to be close to
what is expected from the 22\,GHz image.  The flux densities of the RA
East, RA Central and RA West are 456\,mJy, 2106\,mJy and 845\,mJy,
respectively in the 43\,GHz CC decomposition of the top panel of
Figure~9. Gaussian fits to the nuclear region of the
  RadioAstron data yield flux densities of the RA East, RA Central and
  RA West of 564\,mJy, 775\,mJy and 389\,mJy, respectively at 22
  GHz. RA Central and RA West have a highly inverted (rising) spectrum
  between 22\,GHz and 43\,GHz either from synchrotron self-absorption
  or from free-free absorption \citep{wal00}.

We try to assess the degree to which the nuclear region has changed in
terms of morphology and size in 2018-2020 compared to 2013. The
cross-identification of components is rather uncertain due to 5 years
between the observations. We cannot make any robust
  cross-identifications of the components from epoch to epoch due to
  the known variability of the morphology and the effects of blending
  with low resolution. We simply give two plausible
  cross-identifications in the bottom panel of Figure~9 and Figure~10
  to help elucidate the large qualitative changes that might be
  occurring over time. First, we identify the brightest component of
  VLBA with the brightest component of RadioAstron. With this chosen
  scenario, we tentatively identify RA Central as the same physical
feature as the WN in Figure~4. We also overlay the locations of the EN
in August 26, 2018 and May 12, 2019 on the RadioAstron image in the
bottom panel of Figure~9. In this cross-identification scenario, the
nuclear structure has widened by $\sim 40\mu\rm{as}$ and the whole
pattern is shifted to the east.

\begin{figure}
\begin{center}
\includegraphics[trim={5cm 0 0 0}, clip, width=110mm, angle=0]{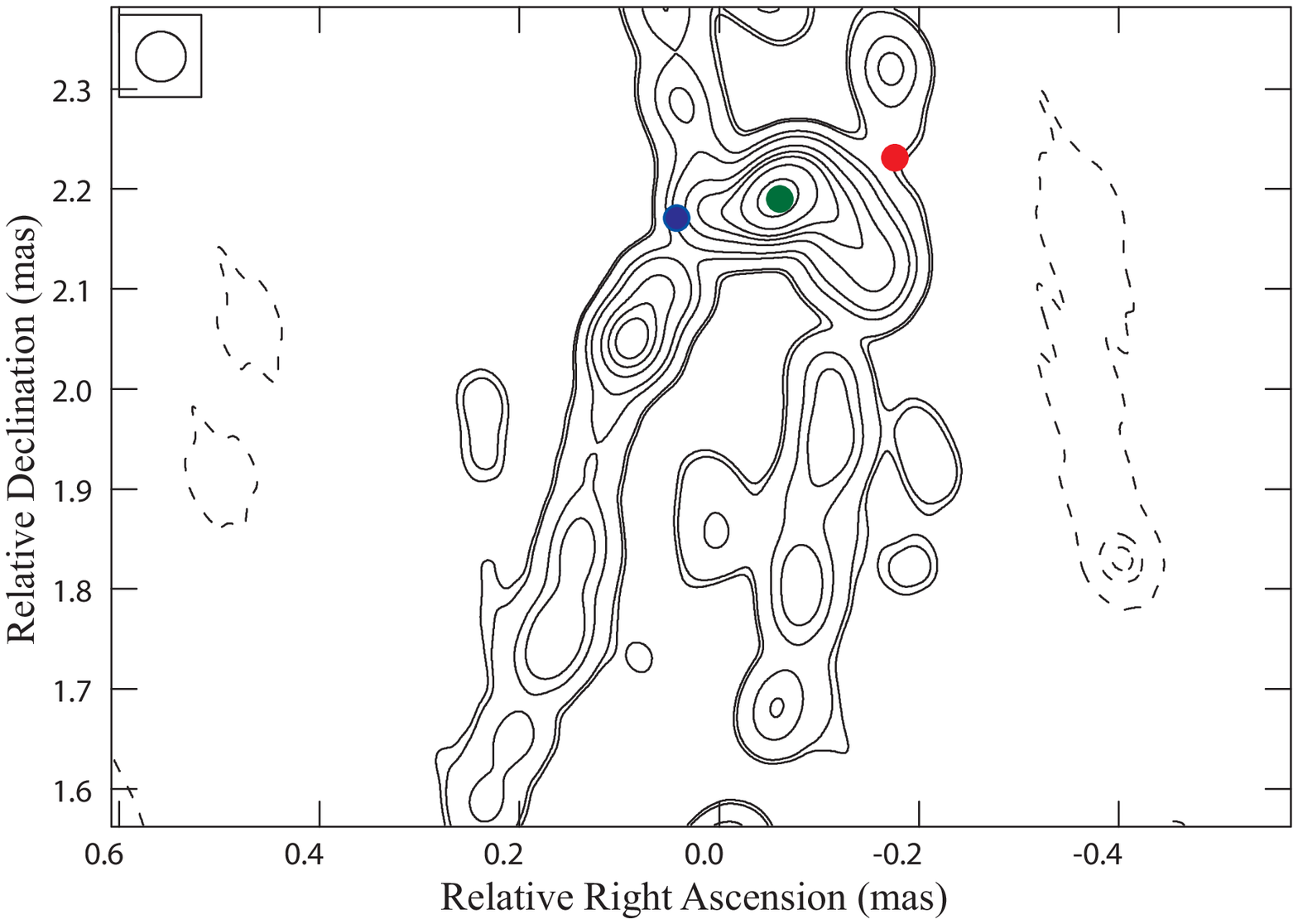}
\caption{We consider an alternative (to Figure~9) cross-identification
  of components between our 43 GHz VLBA monitoring of the multiple
  nucleus and the RadioAstron image. This interpretation is based on
  the CC model for January 4, 2020 in the bottom right hand panel of
  Figure~4. There is a triple nucleus in that epoch as well. The color
  coding of the centroid locations are the same as in Figure~4. One
  very significant difference is that in January 2020 the cental
  component is the weakest and in September 2013 it is the
  brightest. The multiple nucleus is wider in 2020. The contour levels
  are (-10, -7.50, -3, 7.500, 10, 30, 50, 100, 150, 200, 300,
  500)mJy.}
\end{center}
\end{figure}

The first cross-identification scenario is straightforward,
  but just one of many possible choices. Thus, in Figure~10 we show a
different plausible cross-identification based on the high SNR January
4, 2020 observation of the nuclear triple. The color coding of the
triangles corresponds to the color coding in the CC model in
the bottom right hand panel of Figure~4. Note that the
triangles have a different identification than in
Figure~7. In this scenario the nuclear structure has widened by $\sim
60\mu\rm{as}$ and the point of origin of the ejections is the red
triangle at the far west edge of the jet. The triple axis is
tilted relative to September 22, 2013. In summary, both the
interpretation in Figure~9 and Figure~10, indicate a highly dynamic
nuclear region that evolved to a wide state in 2019. The physical
identification of the components is not obvious, but our time
evolution study clearly identified luminous plasma ejected towards the
east.

\section{Conclusion}

In this paper, we consider three methods for resolving the nuclear
structure in 43\,GHz VLBA observations of 3C\,84, intensity
cross-section, Gaussian fitting in the $(u,v)$ plane and CC
models. The time averaged separation speed from August 2018 to April
2020 is estimated as $0.086\pm 0.008$\,c. A second ejected component
was identified in October 2019 to April 2020 with a similar speed.
The multiple nucleus has been detected before with RadioAstron and
86\,GHz global VLBI, but the dynamical evolution has never been seen
previously. This separation speed estimate accomplishes the primary
goal of this paper.

The most relevant comparison are the component measurements at the
base (inner $\sim 0.2$\,mas) of the 43\,GHz VLBA southerly directed
jet \citep{dha98}. They found apparent velocities in the range
0.035\,c $-$ 0.1\,c. However, there was uncertainty in the
identification of the location of nucleus. Independent of any of the
nuclear identification scenarios, the apparent velocity was slow by
relativistic standards $\lesssim 0.1$\,c, very similar to what was
found in this study of the multiple east-west nucleus. The apparent
jet speed accelerates as the jet propagates southward with $\sim
0.25$\,c at $\sim 1$\,mas and the highest estimates approach 0.47\,c
$\sim 2$\,mas away \citep{suz12}.

It is tempting to try to understand the observed motion in the context
of the complicated circum-nuclear structure depicted in
Figure~1. However, we choose not to speculate and use our result as
one piece of the puzzle that must be incorporated into future efforts
to explain the base of the jet in 3C\,84. It is certainly odd to see
an inner flow or expansion that appears to be orthogonal to the large
scale jet that is clearly delineated only $150$\,$\mu$as downstream
\citep{gio18,kim19}. One thing that we can say is that the double
separation is $\sim 0.3$\,lt-yrs on the sky plane in April 2020. This
corresponds to an intrinsic distance of $\approx
0.3/\sin{\theta}$\,lt-yrs, where $\theta$ is the line of sight (LOS)
angle to the component motion. This equates to $\sim
(1900/\sin{\theta}) M$ where $M\sim 10^{9}M_{\odot}\sim 1.5\times
10^{14}\rm{cm}$ is the black hole mass in geometrized units
\citep{nag19,pun18,sch13}. Thus, this slow apparent separation speed
persists very far from the source. The LOS angle has been estimated at
$11^{\circ} - 65^{\circ}$ based on jet/counter-jet asymmetry
\citep{wal94,asa06,lis09,fuj17}. However, these viewing angle
estimates are for the north-south jet. The direction of the east-west
core motion may differ significantly from this. It is perhaps more
relevant that historical data indicates that 3C\,84 sometimes appears
moderately blazar-like based on broad band variability and optical
polarization as high as 6\% \citep{ver78,ang80,chu85,nes95}. In
current nomenclature, it is a slightly ``off angle blazar" at times
\citep{pun18}. For the sake of comparison, estimates for blazars are
typically $\theta \approx 1^\circ - 3^{\circ}$
\citep{hov09,jor17}. However, the blazar-like properties have been far
more benign for the last 35 years \citep{pun18}. As an example, in the
last 12 years the optical polarization has been consistently between
1\%-3\% with the very rare jumps to
4\% \footnote{http://www.bu.edu/blazars/VLBAproject.html}. So we
cannot rule out $\theta \sim 10^{\circ}$, but this behavior does not
favor an extreme blazar-like line of sight. Based on the published
ranges of $\theta$, the physical distance between the EN and WN is
$\sim 0.3- 1.5$ lt-yrs, or $\sim 1900M - 9800$ M. What is extremely
odd is that at this large distance, the VLBA images seem to indicate a
dramatic change in the jet direction. The most straightforward
interpretation is that we are detecting the internal dynamics of a jet
that is $\sim 2000$\,M wide. The lateral expansion near the central
engine proceeds sub-relativistically at $\sim 0.086$c. The jet is
collimated on scales $> 2000$\,M. The RadioAstron image and the April
2020 image in Figure 2 seem to indicate a continuous flow between the
central, wide core component and bright components moving along the
edges of the jet going south.  \par Based on RadioAstron and this
analysis, the bifurcating nuclei are bright ejections of plasma that
can go primarily to the east or west. In the epochs considered here
there is an ejection to the east. In 2020, the images in Figure 2
appear to show that the EN trajectory starts to bend toward the
south. It might be joining the highly collimated portion of the jet
that is directed almost due south. However, this is not so simple
based on Figure~2 because the western ridge is much brighter at this
time. The ejection might traverse (twist around) the front face of the
southerly directed jet and join the western ridge.

The discussion above shows how confusing the dynamics of the inner jet
are likely to be. Considering the variations in the ridge brightness
distribution over time, suggests that the jet dynamics are not even
consistent from epoch to epoch. A year long series of target of
opportunity, 43\,GHz relative astrometry VLBA observations of
$>8$\,hours with two calibrators triggered by a $>0.15$\,mas
separation of the nuclear double could be a feasible method of
tracking the time evolution of the ejections accurately. The
observations could be triggered by a separation determined by the
methods of this paper from the BU data. Perhaps 86\,GHz observations
with global VLBI (including the Atacama Large Millimeter Array) spread
out over 6 months that are triggered simultaneous with the VLBA
observations might shed light on what kind of dynamics are occurring.
These are challenging observations, but this might be one of our best
laboratories for studying the base of a powerful extragalactic jet.

\begin{acknowledgments}
This study makes use of 43\,GHz VLBA data from the VLBA-BU Blazar
Monitoring Program VLBA-BU-BLAZAR funded by NASA through the Fermi
Guest Investigator
Program\footnote{http://www.bu.edu/blazars/VLBAproject.html}. The Very Long Baseline Array (VLBA) is an instrument of the National Radio Astronomy Observatory. The National Radio Astronomy Observatory is a facility of the National Science Foundation operated by Associated Universities, Inc. TS was supported by the Academy of
Finland projects 274477 and 315721. HN is supported by JSPS KAKENHI grant No. JP18K03709.

\end{acknowledgments}

\appendix
\section{The Resolution Limits of the VLBA Observations}

The ability to resolve structures in interferometric data depends on
several factors, e.g., the length of the longest baselines, gaps in the $(u,v)$ coverage, signal-to-noise ratio of the
  data, and the complexity of the source structure itself. Giving an accurate number for any given experiment
requires simulations of that exact experimental setup and the
source. Therefore, one usually resorts to simple rules of thumb
like the fringe spacing, the uniformly weighted beam size or the
criterion we have adopted for the current paper. Since all the
observations we consider have similarly high SNR and similar
  complexity of the source structure, it is reasonable to assume that
any variation in the resolution limit from epoch to epoch is mainly
due to the longest baselines available. Since we use only epochs that
have either MK-SC baseline or MK-HN baseline present, the variation in
the maximum baseline length in E-W direction is only $\sim
15\%$. The important question is, whether our adopted resolution limit
of $0.08 - 0.10$\,mas is on average correct for the VLBA in the high
SNR case. We explore this by means of simulations in the visibility
domain.

\begin{figure}[htp!]
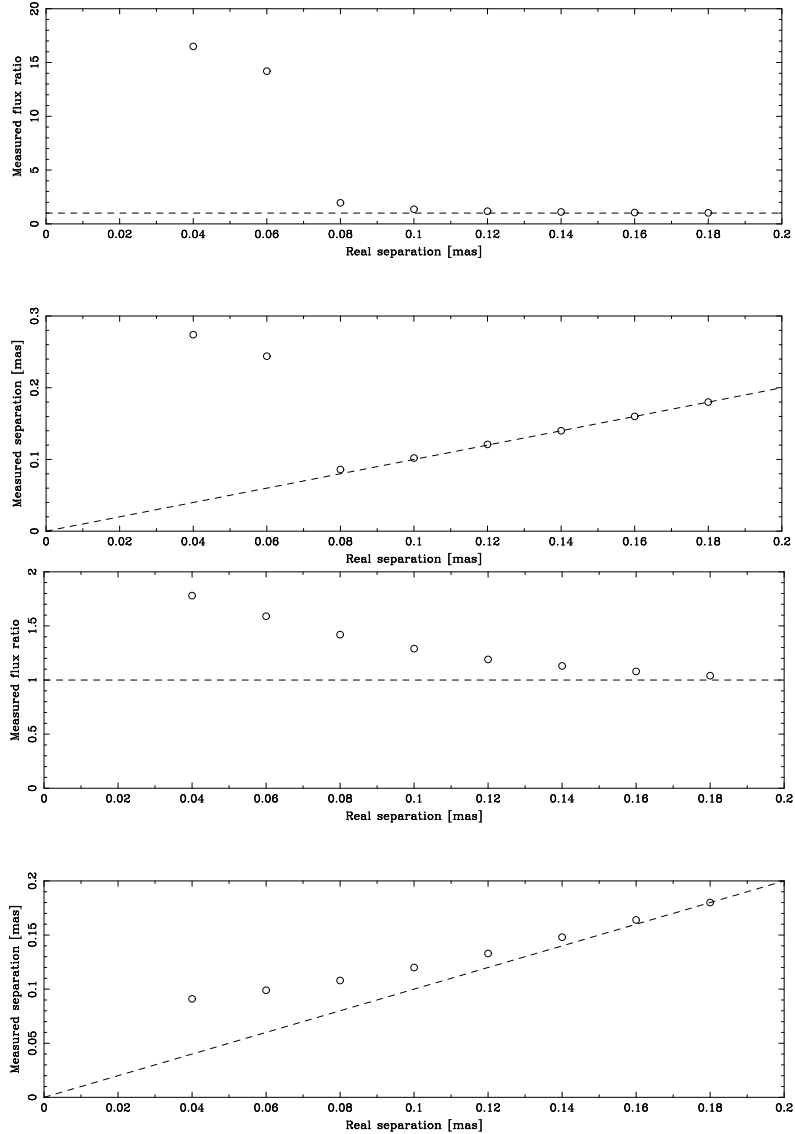

\begin{center}
\includegraphics[width=75 mm, angle= -90]{f44.eps}
\includegraphics[width=75 mm, angle= -90]{f45.eps}
\caption{The results of our simulations in the $(u,v)$ plane. The
  horizontal axes are the separation between our two injected
  nuclear components. The vertical axes are the model-fitted
  values of the flux ratio (for which the ground truth is one by
  construction) and the separation of the two fitted components. The
  top two panels are \ts{for} the point source experiment described in
  the text. Accurate model fitting occurs for a separation of
  $>0.08$\,mas. The bottom two frames are for Gaussian sources
  with a FWHM of 0.1\,mas. In this case a slightly larger separation
  is required for accurate model-fitting in the $(u,v)$ plane.}
\end{center}
\end{figure}

Using \textsc{aips} task \textsc{uvmod} we simulated visibility
  data with exactly the same $(u,v)$ coverage and noise properties as
  in the August 2018 observation, the epoch of minimum detectable
  separation with our CC based method presented in this paper. As an
  input model, we used the August 2018 CC model. From this CC
  model, we excised an area of 3 uniform beam sizes (FWHM) around
  the core. To the resulting model, we added two 1\,Jy point
  sources separated from the origin symmetrically in relative RA
  by $\pm 0.02$\,mas. The independent variable in this experiment is
  the separation in relative RA, the horizontal axis in Figure~11,
  measured in mas. We created eight simulated data sets increasing the
  separation between the point sources from 0.04\,mas to
  0.18\,mas in steps of 0.02\,mas. Then we used the \textsc{modelfit} task in Difmap to fit the visibility data with a model consisting of two point sources in the core region. After making the fit, we plotted the two
  dependent variables of the experiment, the flux ratio and the fitted
  separation of the components in Figure~11, the vertical axis in
  the top and second panels, respectively. At separations of 0.04\,mas
  and 0.06\,mas, the fitted data do not accurately represent the real
  separation nor the real brightness ratios. At 0.08\,mas, this
  abruptly changes to precise fits to both quantities. It clearly
  shows that two equally bright point sources can be resolved, if
  their separation is ~0.08\,mas or larger.

We also made a second set of simulations with the same input
model. However, instead of point sources we used two 1\,Jy Gaussian
distributions with a FWHM of 0.1\,mas each. Recall from Section 3.1,
there were 20 estimates of the FWHM of the components that range from
0.023\,mas to 0.112\,mas with a mean of 0.076\,mas. The fitting in
Difmap was made with point sources as in the previous experiment. The
results are presented in the third and the bottom panels of
Figure~11. It seems that separations of $0.10-0.12$\,mas ($\sim 10\%$
accuracy in the separation) or larger are needed in this case to
accurately estimate the relative positions. We note that the smallest
measurable distance with our CC method in Table~2 is $0.116 \pm
0.013$\,mas in August 2018. Not coincidentally, the value obtained by
the simulation. We also note that the analysis of the simulations
used an inexact model of the flux distributions. This analysis seems
consistent with the model-fitting in the $(u,v)$ plane in
Figure~6. When compared to the CC model method, the model fits seem
less accurate with higher model dependence for separations less than
$\sim 0.13$\,mas.

\newpage

\section{The CC Model from January 10, 2019}
The January 10, 2019 intensity profile looks different than the adjacent epochs in Figure 3.
It has one peak instead of two. Thus, this
epoch needs to be studied in detail. From Section 2, we repeat the
fundamental philosophy of implementing CC models ``Even though the
models may not be robust in a single epoch, they should be
suitable for identifying a trend that persists over time, especially
with regards to very bright features as is the case here." We use this
as our guide for assessing what has occurred during this epoch.

\begin{figure}
\begin{center}
\includegraphics[width=120 mm, angle= 0]{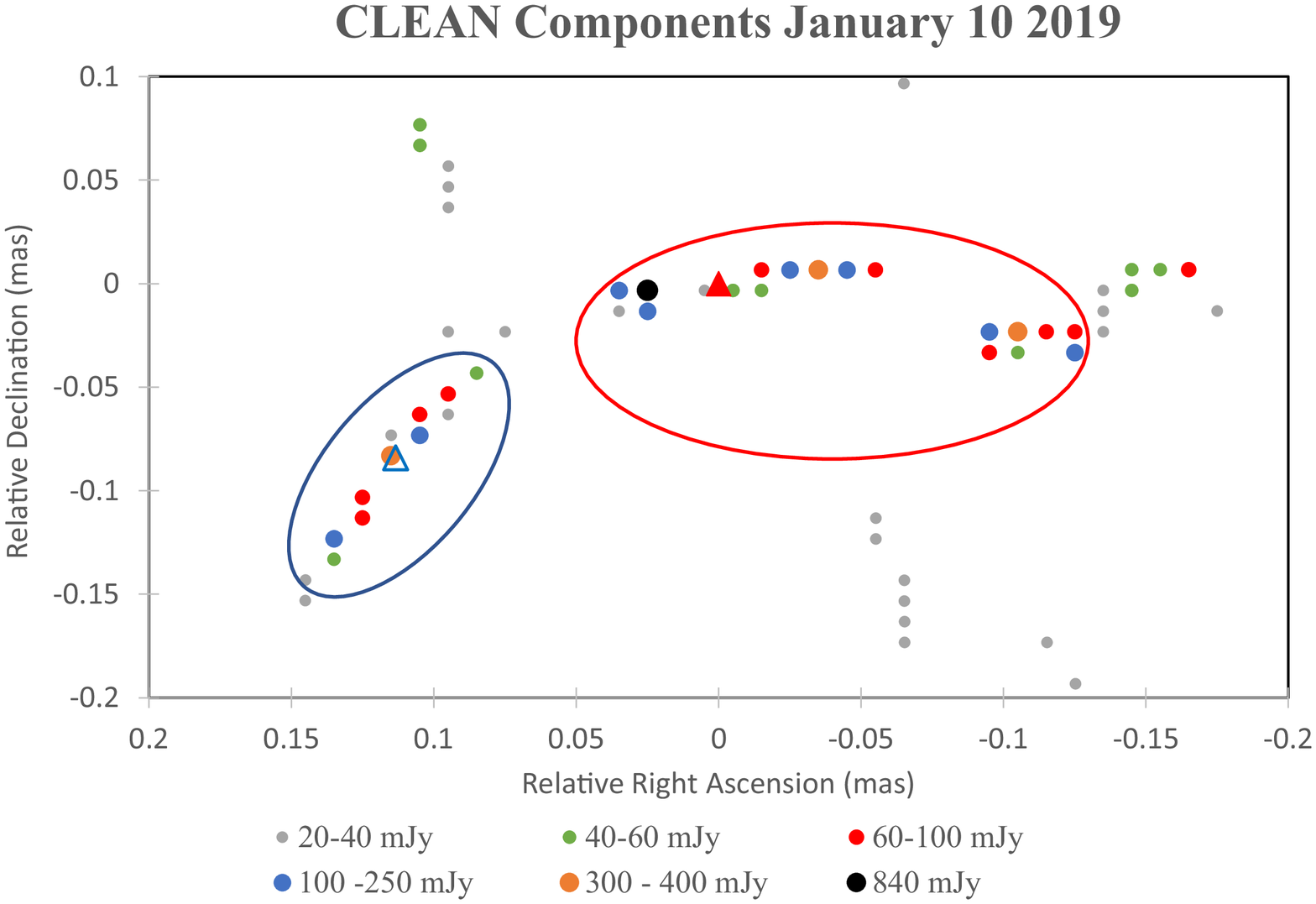}
\includegraphics[width=120 mm, angle= 0]{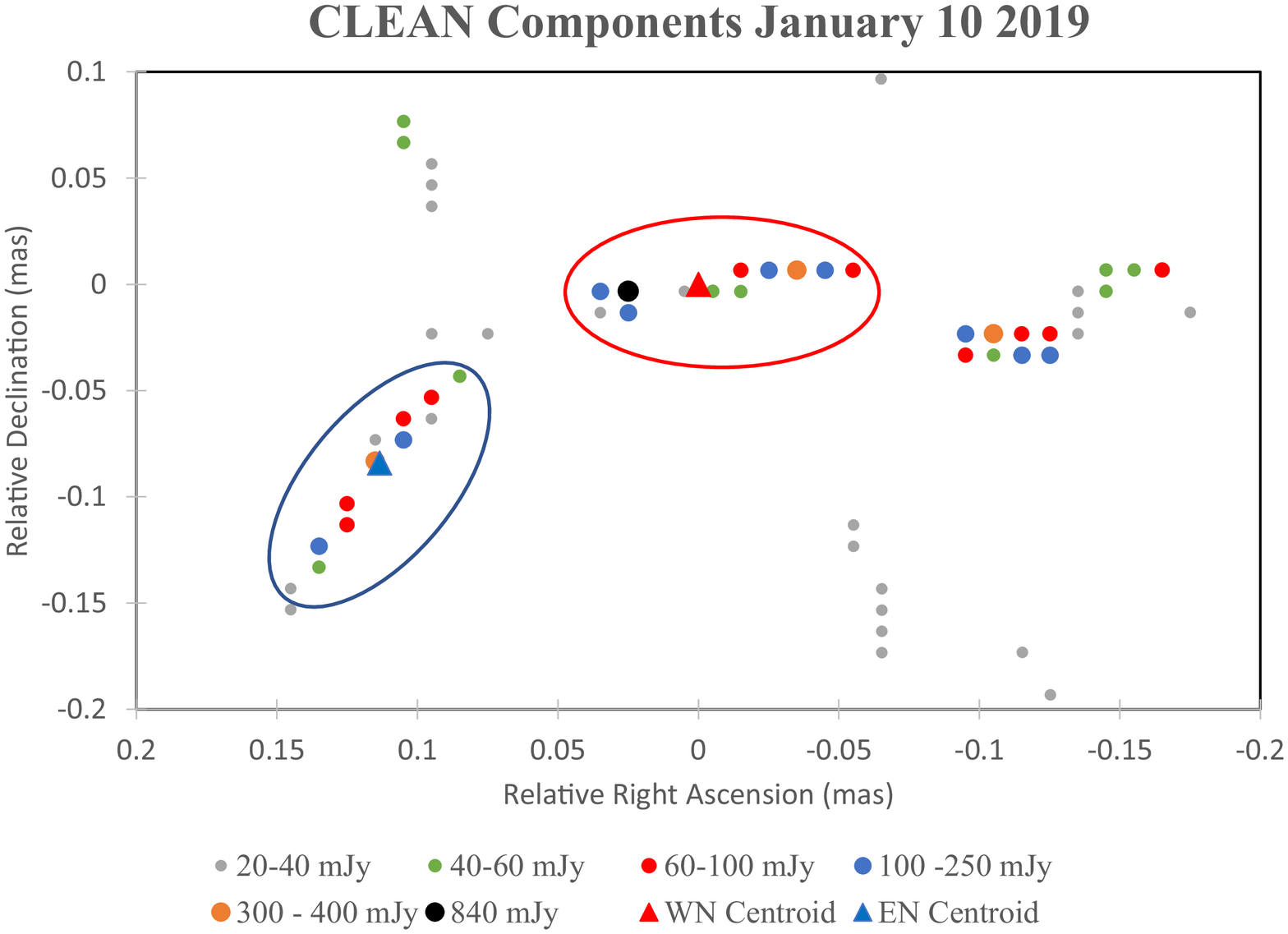}
\caption{The distribution of CCs from the January 10,
    2019 data reduction on the BU website. The figure format is the same as
    Figure 4. Two models are
    presented for the CC clustering. The top one is the one used in
    the main text and Figure 5 and is referred to as Model 1 for the
    clustering of the CCs. Model 2 is the bottom frame.}
\end{center}
\end{figure}

Figure 12 is an analog of Figure 4 that displays the distribution of
CCs. The figure considers two plausible decompositions into a WN and
EN. The single peak structure of the intensity profile is explained by
the figure, the EN lies to the south of the WN by $\lesssim 0.1$
mas. Thus, it is not captured by the intensity cross section that is
centered on the peak intensity (near the centroid of the WN). The
other odd feature is the core now appears to have a bright feature
emitted to the west. This is slightly to the south of the intensity
peak and appears as an inflection point in the western skirt of the
intensity profile in Figure 3. Such a feature would be consistent with
the nuclear triple detected by RadioAstron \citep{gio18}. Our main
task in this Appendix is to investigate if these two features, the
southern displacement of the EN and the western new component, are
accurate reconstructions of the source or affected by artifacts of the
observation and/or the CC model.

The first thing to consider is the quality of the
observations. January 10, 2019 appears to be a good epoch with all 10
antennas operating normally, only with MK and Pie Town (PT) taken out
for the last 2 scans of 3C84 for use by the Navy. Thus, we look for
other indicators of the root cause of the morphological change bearing
in mind the fundamental qualifying assumption of using the CC models
that was restated at the beginning of this Appendix.

Do the other epochs tell us anything? None of the other epochs,
including the one just before in December or the one after in March
have a EN this far south relative to the WN nor do they have an
ejection to the west. The putative ejection to the west is very bight
(1.095 Jy), brighter than the EN, so why does it not appear in March
and May? Such rapid changes could occur with blazar-like
phenomena. However, we are seeing a steady mildly relativistic
separation of the nucleus in Figure 5. If we treat the changes in the
CC model in January as real then it would imply an abrupt change from
this slow (by relativistic standards) steady nuclear separation to one
in which the EN veers south and a bright component is ejected
west. Then in March it appears to be an extrapolation of the slow,
steady separation that was occurring from August to December. The EN
pops back up to the north, similar to previous epochs, and the bright
west ejection disappears and this basic structure persists in
May. This is not impossible (and would be very interesting), but it is
not supported by any other evidence that we have. This would
correspond to the bottom identification of CC clustering for the EN
and WN in Figure 12. We will refer to the bottom frame in Figure 12 as
Model 2, since it is not our preferred model.

The CC clustering model (Model 1) in the top frame of Figure 12 assumes
that the western bright spot is not a new ejection or background
feature, but is separated from the peak intensity possibly as an artifact of the $(u,v)$ coverage. Thus, the two features
are combined into one feature that is stretched out in the east-west
direction in this interpretation. Similarly, the southward veer of the
EN would be a result of imperfect $(u,v)$ coverage and calibration. The
disturbing aspect for our method is that instead of the east-west
separation of the CCs comprising the WN being $\sim 0.1$ mas, it is 0.16 mas. We were motivated
to prefer this interpretation by the smooth transition to and from the
epochs adjacent to it in time. Model 1 is used in Figure 5. This model has a displacement that
is fairly close to the least squares fit in both panels of Figure~5.

Now, we can compare this with Model 2. We
show the fits in Figure 13 that are the analogs of those from Figure 5,
but we have substituted Model 2 for Model 1 in Figure 5. The fits are similar to Figure 5, but Model 2 lies very far from the least squares fit of the east-west displacement. Thus, Figures 5 and 13 favor Model 1 over Model 2 based on the trends of the other epochs.

\begin{figure}
\begin{center}
\includegraphics[width=110 mm, angle= 0]{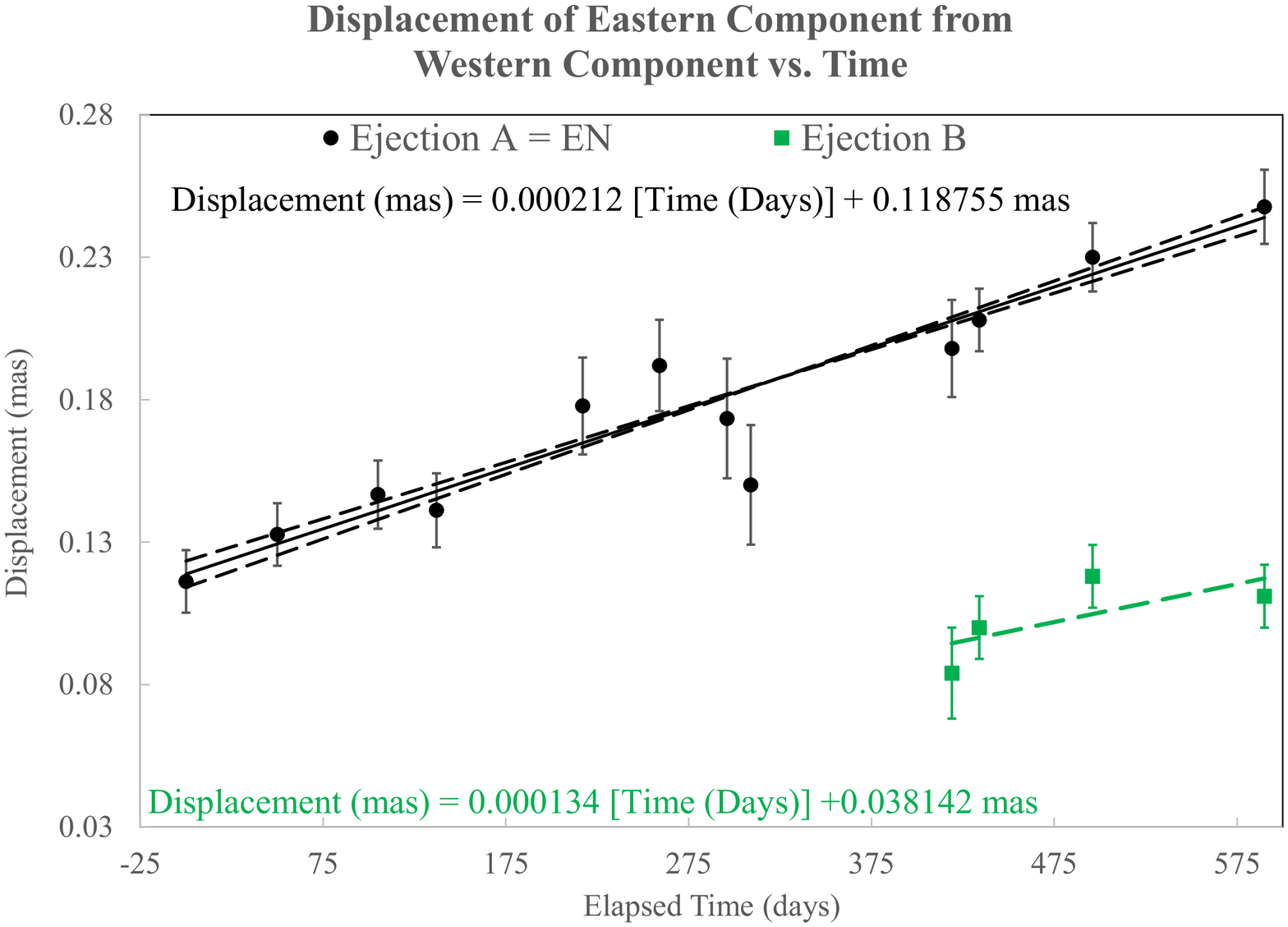}
\includegraphics[width=110 mm, angle= 0]{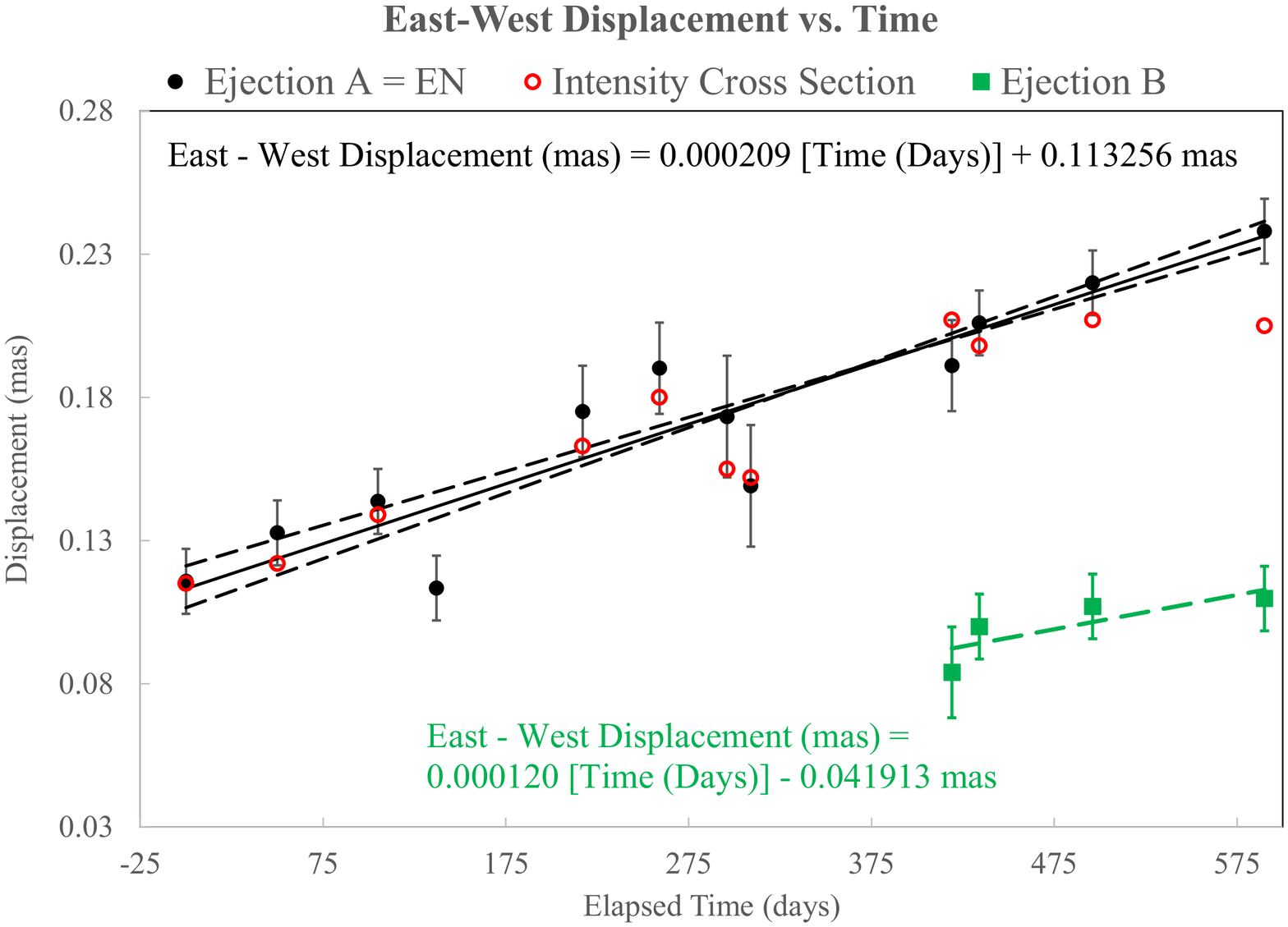}
\caption{Figure 5 is recreated, but instead of using
    Model 1 from the top panel of Figure 12, we use Model 2 from the
    bottom panel of Figure 12 to estimate the double nucleus separation
    speed.}
\end{center}
\end{figure}
\par The anomalous intensity cross-section in Figure 3 and the ambiguity in the component
decomposition (two or three components) of the nucleus is disturbing considering that this
observation had all 10 stations. In order to investigate this further, we tried re-imaging the $(u,v)$ data with CLEAN + self-cal since the end result of the image reconstruction depends on the ``route" that is taken during the imaging process for imperfect data (i.e., sampling is poor and/or data has significant gain errors). The resultant images are sensitive to the initial phase self-cal steps and we were able to redistribute the CCs in the image plane. However, none of the images had an intensity cross section that was ``in-family" with the cross section expected from Figure 3 (i.e., 2 well defined, distinct peaks). Furthermore, none of the images resolved the ambiguity between a two component decomposition of the nucleus or a three component decomposition of the nucleus. Similar to Figure 12, the EN-WN separation was always in the range 0.13 mas to 0.17 mas, regardless of the self calibration or the particular choice of the identifications of the CCs with the EN and the WN. It is therefore concluded that the ``out of family" properties of the January 2019 observation are inherent to the $(u,v)$ data and are not an artifact of image reconstruction. The re-imaging investigation does not address whether this is a physical property of the source at the time of observation or an idiosyncracy of this particular patchy sampling in the $(u,v)$ plane.
\par We hypothesize, without proof, the plausibility of $(u,v)$ coverage issues
with the observation. 3C84 is a very complex source with many
components. The $(u,v)$ data sampling is not ideal for such a complex
source, as it is being observed with 31-32 other sources over a 24
hour period. The $(u,v)$ coverage is not the same in all
epochs. Occasionally, the $(u,v)$ coverage might not be sampled well
enough to reconstruct the complex structure adequately or uniquely. We
cannot prove if the image is accurate or adversely affected by patchy
$(u,v)$ coverage. Our basic premise is that we cannot vouch for the integrity of the CC models in an individual
epoch, but we can detect temporal trends in the bright components. To this point, the results of our analysis do not depend on
choosing Model 1 over Model 2. The estimated separation velocity with the choice of Model 1 in Figure~5 was $0.086\pm 0.08$\,c. By comparison, using Model 2 (as was done in Figure~13), results in almost exactly the same separation velocity $0.088\pm 0.007$\,c. Thus, our results do not depend on the confusing details of a particular epoch.

\end{document}